\documentclass[fleqn,usenatbib]{mnras}

\usepackage{txfonts}

\usepackage[T1]{fontenc}

\DeclareRobustCommand{\VAN}[3]{#2}
\let\VANthebibliography\thebibliography
\def\thebibliography{\DeclareRobustCommand{\VAN}[3]{##3}\VANthebibliography}

\usepackage[final]{graphicx}	
\usepackage{mwe} 
\usepackage{multirow}
\usepackage{booktabs}
\usepackage[flushleft]{threeparttable}
\usepackage{caption}
\usepackage{subcaption}
\usepackage{comment}
\usepackage{csquotes}
\usepackage{scrextend}
\usepackage{float}
\usepackage{txfonts}
\title[Interstellar N-containing PAHs]{Rotational spectra of interstellar N- and CN-PAHs: pyrene and coronene}
\author[Vats et al.]{
Akant Vats\thanks{akant.vats@bhu.ac.in} and
Amit Pathak\thanks{amitpah@gmail.com}
\\
Department of Physics, Institute of Science, Banaras Hindu University, Varanasi-221005, India\\
}

\date{Accepted XXX. Received YYY; in original form ZZZ}

\pubyear{2021}

\begin{document}
\label{firstpage}
\pagerange{\pageref{firstpage}--\pageref{lastpage}}
\maketitle
\begin{abstract}
The detection of benzonitrile (C$_{6}$H$_{5}$CN), 1- and 2-cyano-naphthalene (C$_{10}$H$_{7}$CN) in the cold, dark molecular cloud TMC-1 at centimetre (cm) wavelengths has opened up prospects for the detection of other N- and CN-containing polycyclic aromatic hydrocarbons (PAHs). In this light, the pure rotational spectra of N-pyrene (C$_{15}$H$_{9}$N), CN-pyrene (C$_{15}$H$_{9}$CN), N-coronene (C$_{23}$H$_{11}$N) and CN-coronene (C$_{23}$H$_{11}$CN) are reported here for the first time. The B3LYP/6-311+G(d,p) level of theory, in the Density Functional Theory (DFT) calculations, achieves the best performance for calculating the spectroscopic parameters and simulating the rotational spectra. The large permanent dipole moment of CN-PAHs makes them the most suitable PAH species for detection in the interstellar medium. Additionally, pyrene's smaller partition function makes CN-pyrene a prime candidate to be discovered in cold, dark molecular clouds such as the TMC-1. The present work sets a benchmark for theoretical rotational spectra of N- and CN-containing PAHs and may act as a guide for laboratory experiments and observational searches.              
\end{abstract}




\section{Introduction}
The mid-infrared (IR) spectra of the interstellar medium (ISM) of our Galaxy and several other galaxies are dominated by the emission features peaking strongly at 3.3, 6.2, 7.7, 8.6, 11.2 and 12.7 $\mu$m, commonly known as unidentified infrared (UIR) bands \citep{Gil73,Coh86,1996PASJ...48L..59O,1999ApJ...513L..65S,2002A&A...390.1089P,2020NatAs...4..339L}. The UIR bands are ubiquitous characteristics of a wide variety of astronomical sources including H~$\textsc{ii}$ regions, post-AGB stars, young stellar objects (YSOs), planetary nebulae (PNe), reflection nebulae (RNe), the ISM of galaxies as well as the diffuse ISM \citep{2001A&A...370.1030H,2001A&A...372..981V,2002A&A...390.1089P,2007ApJ...656..770S,2007ApJ...664.1144S}. These bands are a result of the IR fluorescence involving the vibrational bands of polycyclic aromatic hydrocarbon (PAH) molecules after absorption of ultraviolet (UV) photons \citep{1984A&A...137L...5L,1985ApJ...290L..25A,1989ApJS...71..733A,2004ASPC..309..163O,2008ARA&A..46..289T}. About 15$\%$ of the interstellar carbon may be associated with PAHs and related species and these PAHs affect the chemical and physical processes in their habitat \citep{2008ARA&A..46..289T,2020NatAs...4..339L}. Other than their vibrational emission, the rotational emission from PAHs seems most appropriate carrier of \enquote{anamalous microwave emission} (AME) present in the frequency range of 10--60 GHz \citep{1996ApJ...470..653K,1997ApJ...486L..23L,Hensley_2022}. Furthermore, the electronic spectra of PAHs may interpret some of the diffuse interstellar bands (DIBs); absorption bands observed in the UV and optical \citep{2011ApJ...728..154S}. 
\begin{figure*}
     \centering
     \begin{subfigure}[b]{0.6\textwidth}
         \centering
         \includegraphics[width=\textwidth]{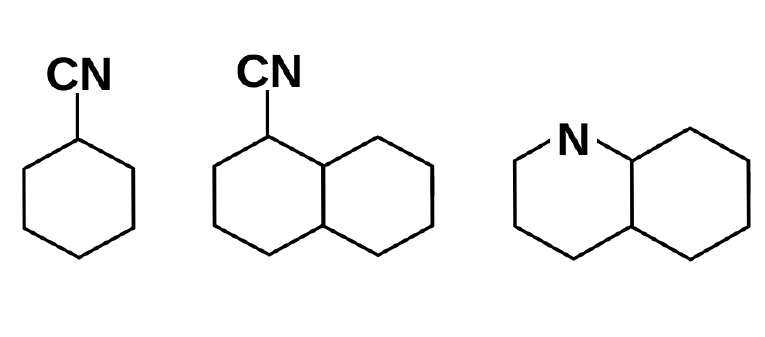}
         \caption{(a) Benzonitrile, 1-cyano-naphthalene and quinoline (from left to right).}
     \end{subfigure}\\
     \begin{subfigure}[b]{1\textwidth}
         \centering
         \includegraphics[width=\textwidth]{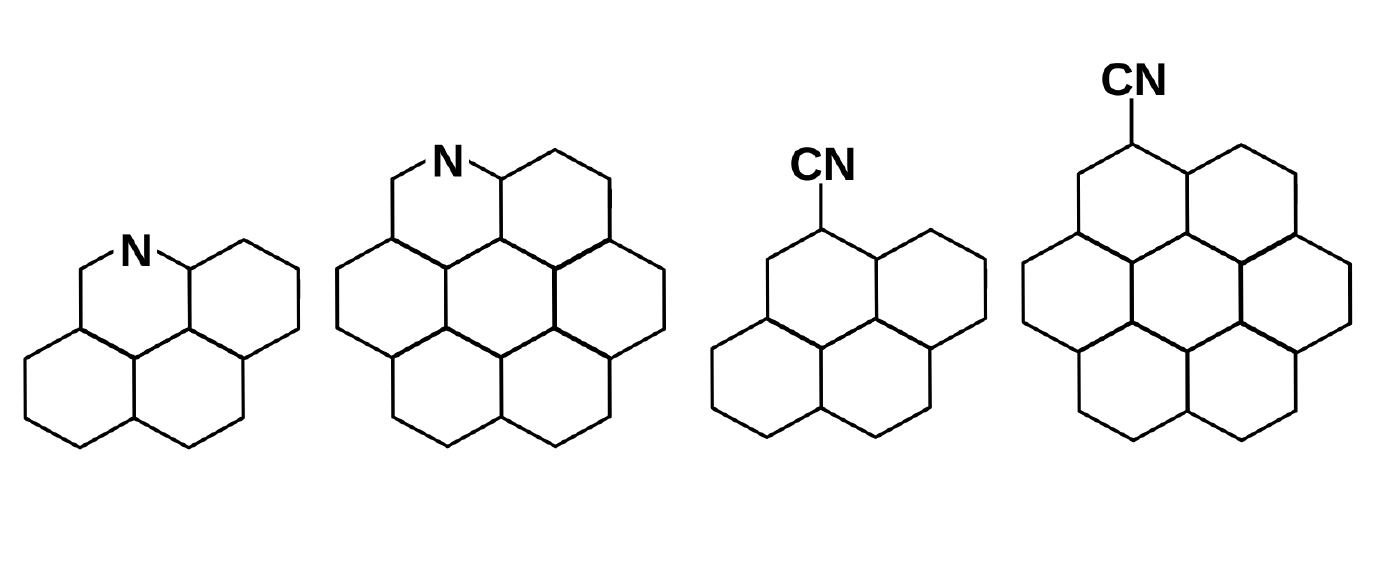}
         \caption{(b) N-pyrene, N-coronene, CN-pyrene and CN-coronene (from left to right).}
         \label{fig:three sin x}
     \end{subfigure}
     \caption{N-containing species considered here as (a) test molecules and (b) target molecules.}
\end{figure*}

Despite the importance of PAHs as a common and ubiquitous interstellar molecule, the detection of a specific PAH in the ISM through UIR bands is quite challenging as they arise from nearest neighbour vibrations and could be mixtures of different PAH species \citep{2008ARA&A..46..289T}. On the other hand, pure rotational spectroscopy is a powerful technique in detection of the interstellar molecules and have been used to identify around 80$\%$ of molecules observed in the space \citep{2018Sci...359..202M}. Due to the lack of a permanent dipole moment in PAHs and structural similarities, this technique is not useful for PAHs. However, a bowl-shaped PAH molecule having a large permanent dipole moment, corranulene (C$_{20}$H$_{10}$) has been searched for using rotational spectroscopy, with no success \citep{2009MNRAS.397.1053P}. Later on, it has been found that planar molecules such as PAHs can have a permanenet dipole moment if they carry some impurity, mainly in the form of substitution or having different side groups \citep{2014MNRAS.437.2728A}. This marks a silver-lining about the brighter prospects of rotational spectroscopy for interstellar PAHs.

Based on the above, other readily detectable PAH species have been studied, out of which the most prominent and suitable PAH candidates are nitrogen containing PAHs or polycyclic aromatic nitrogen heterocycles (PANHs) \citep{2005ApJ...632..316H}. PANHs are strongly polar in nature, likely widespread in the ISM and possible carriers of the UIR band at 6.2 $\mu$m \citep{2005ApJ...632..316H,2021ApJ...923..202R,2022PASJ...74..161V}. PANHs have been detected in carbonaceous meteorites \citep{2002NPRep..19..292S} and in the form of PAH nitriles, such as benzonitrile (C$_{6}$H$_{5}$CN), 1-cyano-1,
3-cyclopentadiene (C$_{5}$H$_{6}$CN), and 1- and 2-cyano-naphthalene (C$_{11}$H$_{7}$CN) in the TMC-1, a cold, dark molecular cloud \citep{2018Sci...359..202M,2021NatAs...5..176M,2021Sci...371.1265M}. This confirms a significant presence of PANHs in the ISM.

Recent technological developments in single dish and interferometry radio antennas have resulted in facilities with great sensitivity and resolution (such as the Atacama Large Millimeter Array (ALMA)), driving up demand for more precise rotational spectroscopic data. 
However, experimental studies regarding pure rotational spectra of PANHs are limited in number and restricted to small PAH species (within 12 C atoms). Therefore, theoretical methodologies might be more convenient to study PANHs. The complete interpretation of AME requires pure rotational spectroscopic data on PAHs/PANHs and other dust-correlated carbonaceous species. 

In the present work, the pure rotational spectra of PANHs are computed and analyzed theoretically in the frequency range of 8 to 33 GHz, relevant to the interstellar search of aromatic molecules \citep{2015MNRAS.447..315A,2018Sci...359..202M,2021Sci...371.1265M,2021NatAs...5..176M}. The accuracy of the computational methods adopted in this work has been verified before applying on the target PANH molecules. The main aim of this work is to provide the pure rotational data of large PANH molecules to explore their possibility of detection in the ISM. The paper is organized as follows. Section 2 describes the selection of the sample and theoretical methods. Section 3.1 presents the effect of different basis sets on rotational constants and establishes the accuracy of the calculation methods used. The results are discussed in Section 3.2. Section 4 discusses the astrophysical implications with conclusions. 
\begin{table*}
\begin{threeparttable}
    \centering
        \caption{\small Rotational spectral parameters of benzonitrile (C$_{6}$H$_{5}$CN), 1-cyanonaphthalene (C$_{10}$H$_{7}$CN) and quinoline (C$_{9}$H$_{7}$CN). The deviation$^{\Pi}$ between the experimental and theoretical values are given in parentheses.}
    \begin{tabular}{lcccc}
    \hline
    \hline
   Constants & Experimental & B3LYP & B3LYP & MP2\\
    &  & 6-311+G(d,p) & cc-pVTZ & 6-31G(d,p) \\\hline
  \hspace{-0.15cm}C$_{6}$H$_{5}$CN$^{\dagger}$ & & & & \\
 A$_{\rm 0}^{\rm a}$ (MHz) & 5655.2652 & 5673.8318 (0.32) & 5703.7911 (0.85) & 5657.1971 (0.03)\\     
 B$_{\rm 0}$ (MHz) & 1546.8758 & 1548.3016 (0.09) & 1555.2297 (0.53) & 1535.6142 (0.73)\\
 C$_{\rm 0}$ (MHz) & 1214.4040 & 1216.3723 (0.16) & 1222.0247 (0.62) & 1207.7712 (0.54)\\
 $\Delta_{\rm J}^{\rm b}$ (Hz) & 45.6 & 43.2 & 43.4 & 41.9\\ 
 $\Delta_{\rm JK}$ (Hz) & 933.1 & 865.6 & 898.2 & 924.6\\ 
 $\Delta_{\rm K}$ (Hz) & 272 & 285 & 263 & 196\\ 
 $\delta_{\rm J}$ (Hz) & 11.10 & 10.43 & 10.53 & 10.20\\ 
 $\delta_{\rm K}$ (Hz) & 613 & 561 & 578 & 590\\
 $\chi_{xx}^{\rm c}$ (MHz) & -4.237 & -4.04 & -4.41 & -3.98\\
 $\chi_{yy}$ (MHz) & 2.289 & 2.19 & 2.38 & 2.17\\
 $\chi_{zz}$ (MHz) & 1.949 & 1.85 & 2.03 & 1.81\\
  $\mu^{\rm d}$ (Debye) & $\mu_{\rm a}$ = 4.5152 & $\mu_{\rm a}$ = 4.7426 & $\mu_{\rm a}$ = 4.6447 & $\mu_{\rm a}$ = 5.0092\\
\hline
\hspace{-0.15cm}C$_{10}$H$_{7}$CN$^{\ddagger}$ & & & & \\
 A$_{\rm 0}^{\rm a}$ (MHz) & 1478.8483 & 1488.1689 (0.62) & 1493.7911 (1.01) & 1473.1330 (0.38)\\     
 B$_{\rm 0}$ (MHz) & 956.7842 & 950.3583 (0.67) & 960.0357 (0.33) & 954.1701 (0.27)\\
 C$_{\rm 0}$ (MHz) & 580.9889 & 579.9785 (0.10) & 584.4317 (0.58) & 579.0868 (0.32)\\
$\Delta_{\rm J}^{\rm b}$ (Hz) & 39.636 & 37.005 & 37.199 & 40.224\\ 
 $\Delta_{\rm JK}$ (Hz) & -159.63 & -140.66 & -137.32 & -151.57\\ 
 $\Delta_{\rm K}$ (Hz) & 249.40 & 223.72 & 221.93 & 232.48\\ 
  $\delta_{\rm J}$ (Hz) & 17.94 & 16.19 & 16.98  & 17.72\\ 
  $\delta_{\rm K}$ (Hz) & 10.29 & 10.81 & 10.93 & 11.72\\ 
  $\chi_{xx}^{\rm c}$ (MHz) & 1.92 & 1.97 & 2.12 & 1.99\\
 $\chi_{yy}$ (MHz) & 2.98 & 2.50 & 2.60 & 2.71\\
 $\chi_{zz}$ (MHz) & -4.90 & -4.47 & -4.73 & -4.52\\
  $\mu^{\rm d}$ (Debye) &  & $\mu_{\rm a}$ = 3.6616 & $\mu_{\rm a}$ = 3.5632 & $\mu_{\rm a}$ = 3.7782 \\
     &  & $\mu_{\rm b}$ = 3.0372 & $\mu_{\rm b}$ = 2.9645 & $\mu_{\rm b}$ = 3.2877\\
 \hline
 \hspace{-0.15cm}C$_{9}$H$_{7}$N$^{\star}$ & & & & \\
 A$_{\rm 0}^{\rm a}$ (MHz) & 3145.5330 & 3154.9974 (0.29) & 3171.7536 (0.82) & 3139.9102 (0.17)\\     
 B$_{\rm 0}$ (MHz) & 1271.5779 & 1271.2549 (0.02) & 1277.5652 (0.46) & 1271.2024 (0.02)\\
 C$_{\rm 0}$ (MHz) & 905.7394 & 906.1403 (0.04) & 910.7285 (0.54) & 904.8712 (0.09)\\
 $\Delta_{\rm J}^{\rm b}$ (Hz) & 19.110 & 18.389 & 18.533 & 18.687\\ 
 $\Delta_{\rm JK}$ (Hz) & 47.031 & 46.456 & 46.883 & 46.891\\ 
 $\Delta_{\rm K}$ (Hz) & 161.461 & 161.222 & 162.859 & 163.024\\ 
 $\delta_{\rm J}$ (Hz) & 5.662 & 5.448 & 5.492 & 5.501\\ 
 $\delta_{\rm K}$ (Hz) & 60.622 & 58.611 & 59.160 & 60.148\\ 
   $\chi_{xx}^{\rm c}$ (MHz) & -4.684 & -4.682 & -4.925 & -4.302\\
 $\chi_{yy}$ (MHz) & 3.221 & 3.219 & 2.866 & 2.791\\
 $\chi_{zz}$ (MHz) & 1.462 & 1.458 & 1.455 & 1.502\\
  $\mu^{\rm d}$ (Debye) & $\mu_{\rm a}$ = 0.1435 & $\mu_{\rm a}$ = 0.2367 & $\mu_{\rm a}$ = 0.2014 & $\mu_{\rm a}$ = 0.216\\
     & $\mu_{\rm b}$ = 2.0146 & $\mu_{\rm b}$ = 2.1572 & $\mu_{\rm b}$ = 2.0314 & $\mu_{\rm b}$ = 2.012\\
  \hline
 \end{tabular}
 \label{tab:my_label}
 	\begin{tablenotes}
	\item Notes. $^{\Pi}$Relative error = $\frac{\rm exp. - theo.}{\rm theo}$ $\times$ 100.
\item $^{\rm a}$Rotational constants, $^{\rm b}$Quartic centrifugal distortion constants, $^{\rm c}$$^{14}$N quadrupole coupling constants in reference to the principal axis system, $^{\rm d}$Dipole moment.
\item $^{\dagger}$\cite{2018Sci...359..202M}, $^{\ddagger}$\cite{2018MNRAS.476.5268M}, $^{\star}$\cite{2003JMoSp.217..115K}.
\end{tablenotes}
\end{threeparttable}
\end{table*}

\begin{figure*}
	\subfloat{\includegraphics[width=6.3cm]{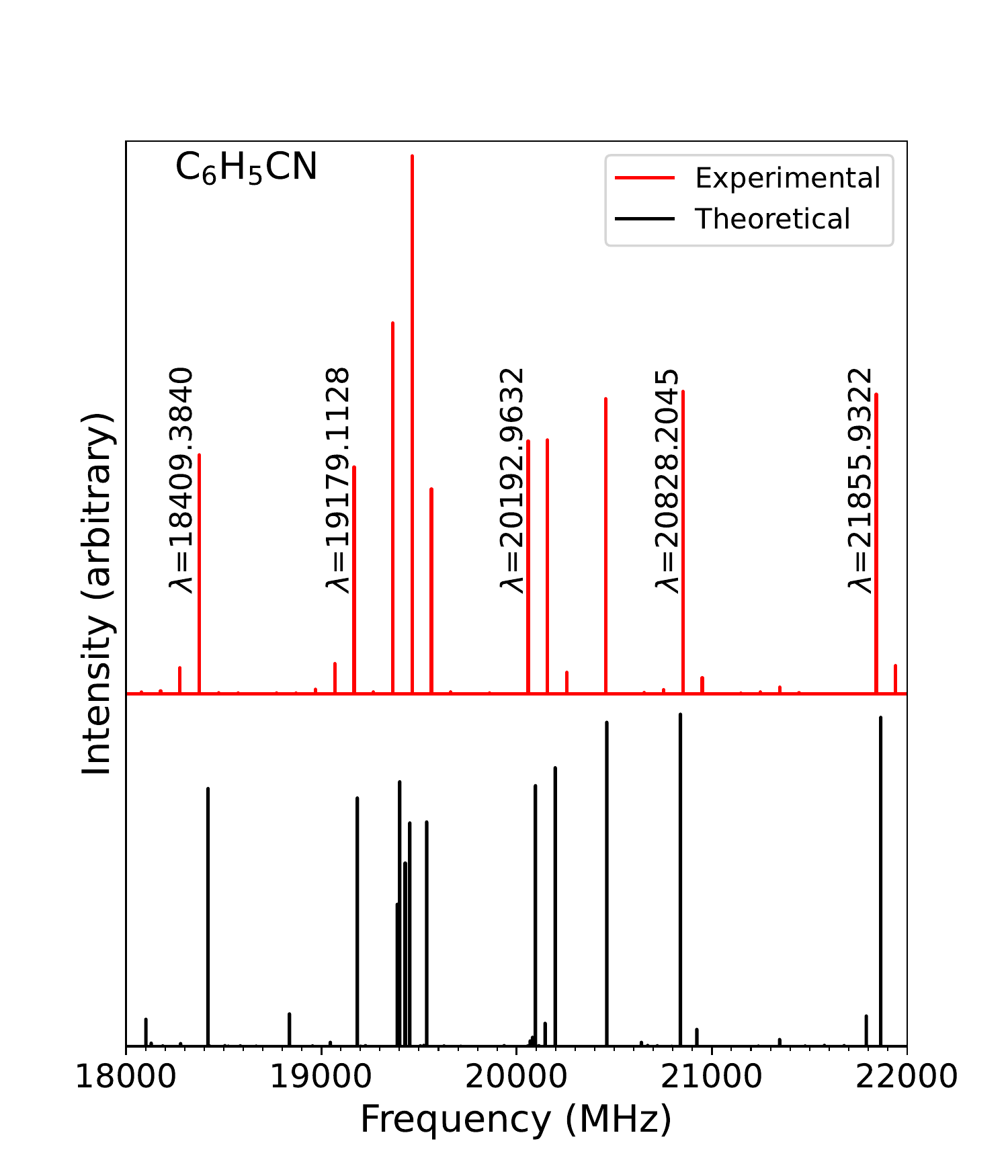}}
\subfloat{\includegraphics[width=6.3cm]{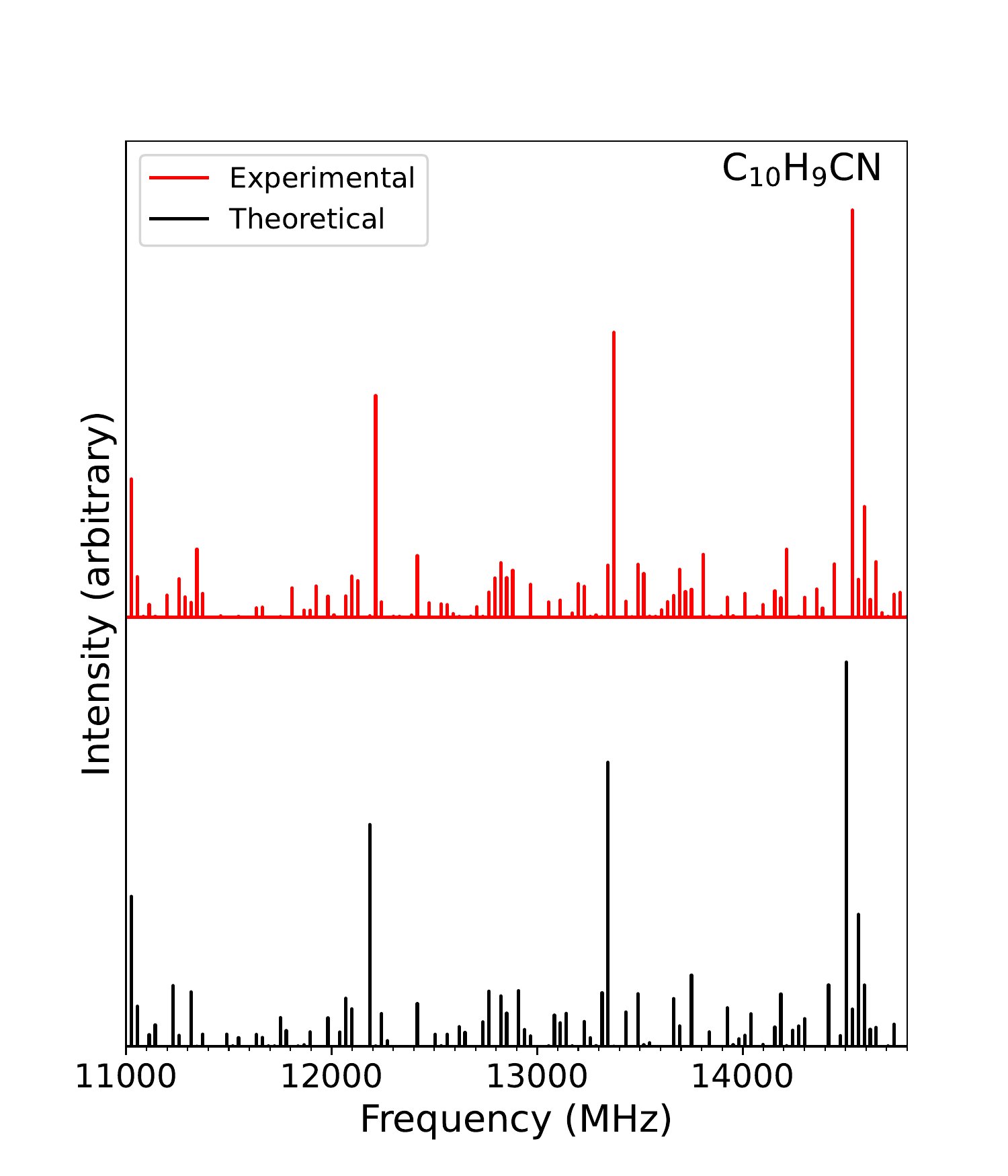}}
	\subfloat{\includegraphics[width=6.3cm]{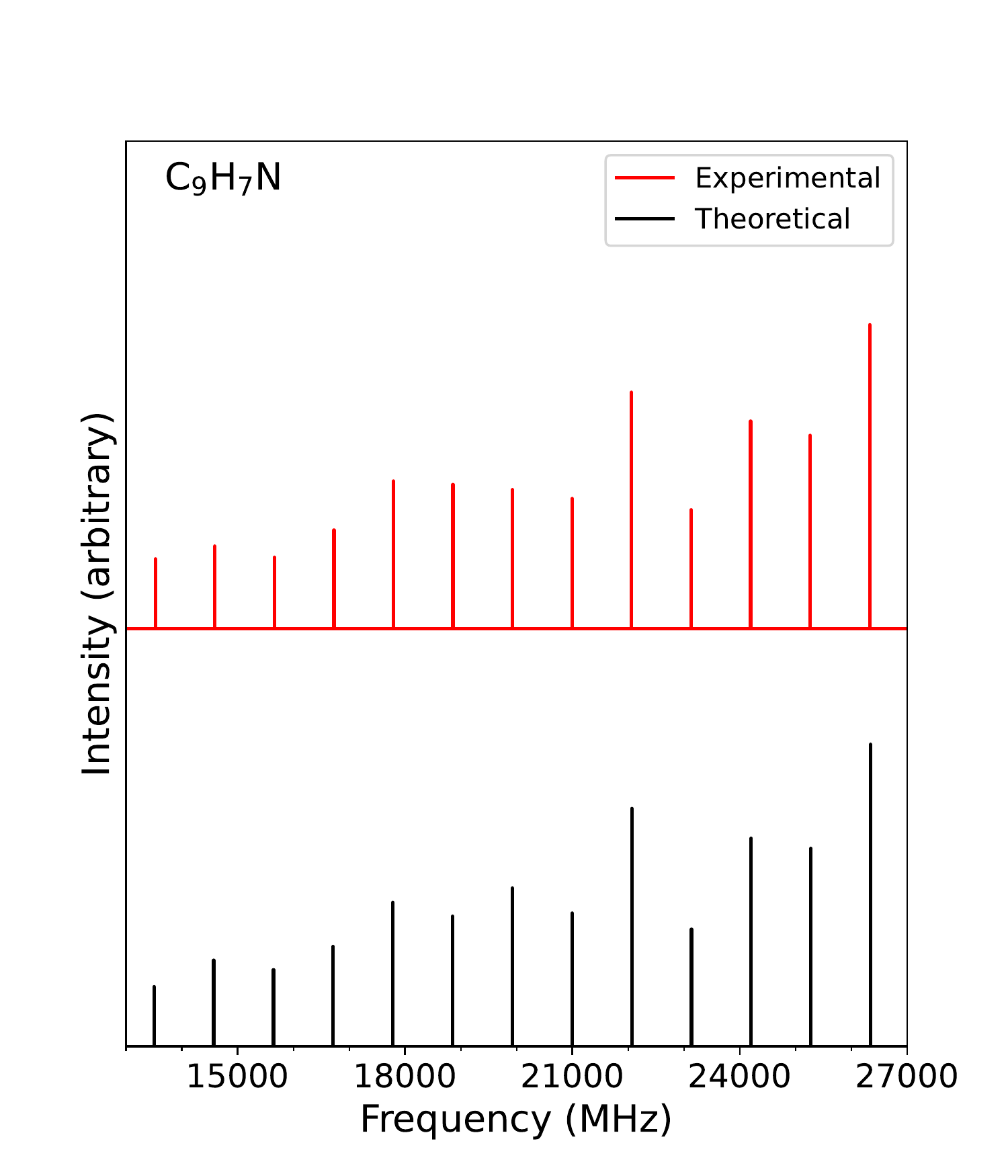}}
       \caption{Rotational spectra of benzonitrile (C$_{6}$H$_{5}$CN), 1-cyanonaphthalene (C$_{10}$H$_{7}$CN) and quinoline (C$_{9}$H$_{7}$N) with B3LYP/6-311+G(d,p) at 300 K (black) in the centimetre-wave region. The experimental spectra are shown in red. The band positions of detected emission lines of benzonitrile in the TMC-1 are labelled.}
    \label{fig:example_figure}
\end{figure*}
\section{Sample and Theoretical Methods}
The comparison of the UIR bands with theoretical IR spectra suggests the abundance of symmetric and compact PAHs in the ISM \citep{2009ApJ...697..311B,2012ApJ...754...75R}. The substitution of N atom in compact PAH forms a quasi-symmetric and strongly polar PANH molecule having well-identifiable rotational spectra \citep{2015MNRAS.447..315A}. Thus, PANHs are suitable candidates to be studied for PAH rotational spectroscopy. There are two most appropriate forms of PANHs--- (i) PAHs having N within the PAH skeleton or N-substituted PAHs (N-PAHs, Fig. 1) \citep{2005ApJ...632..316H,2021ApJ...923..202R,2022PASJ...74..161V} and, (ii) PAH nitriles or CN-attached PAHs (CN-PAHs, Fig. 1). Based on this, we consider nitrogen substitution in two symmetric and compact PAHs [pyrene (C$_{16}$H$_{10}$) and coronene (C$_{24}$H$_{12}$)] having substantial photostability against UV radiation. For N-PAHs, there are two possibilities for N-inclusion in PAHs; exoskeletal (exo-PANHs), where the N atom replaces CH at the periphery, and endoskeletal (endo-PANHs), where N is incorporated within the PAH structure in place of C. There is very little impact on the molecular stability when N is substituted in the exoskeletal manner, however, the incorporation of N atom within the ring causes an excess charge and higher electron repulsion, which reduces stability \citep{2021ApJ...923..202R}. Therefore, the present study considers the N-PAHs, where N is substituted at the periphery. 

Since there are no laboratory rotational studies available for pyrene and coronene in N- and CN-PAH variants, we have selected 3 N-containing molecules for establishing the accuracy of our computational method. These molecules are benzonitrile (C$_{6}$H$_{5}$-CN), 1-cyanonaphthalene (C$_{10}$H$_{7}$-CN), (CN-PAHs) and quinoline (C$_{9}$H$_{7}$N), where N replaces CH at the periphery (N-PAH). The first two have recently been discovered in the ISM \citep{2018Sci...359..202M,2021Sci...371.1265M}, whereas the third has been proposed for identification and has been explored in a number of studies \citep[\& references therein]{2003JMoSp.217..115K,2015JChPh.142j4310P}. 

The GAUSSIAN 09 \citep{Gaussian09} suite of programs was used to compute each molecule's geometry, rotational constants, quadrupole coupling constants, quartic centrifugal distortion constants and dipole moments using Density Functional Theory (DFT) with anharmonic vibrational frequency calculations. DFT has been widely utilized in rotational spectroscopy of interstellar N-containing PAHs to determine the molecular or spectroscopic constants \citep{2003JMoSp.217..115K,2008ApJ...678..309M,2015JChPh.142j4310P,2018MNRAS.476.5268M}, as well as to compute their vibrational modes \citep{2005ApJ...632..316H,2022PASJ...74..161V,2021ApJ...923..202R}. The use of B3LYP functional with the 6–311+G(d,p) basis set or the correlation-consistent polarized triple valence basis set (cc-pVTZ) are found effective for studying the molecular or spectroscopic constants of CN-PAHs \citep{2018MNRAS.476.5268M}, whereas the MP2/6-31G(d,p) combination is found better for quinoline \citep{2003JMoSp.217..115K}. As a result, the current study employs all three of the above-mentioned basis sets on the test molecules (Fig. 1-a). The basis set with the best match is then used to compute the target molecules (Fig. 1-b) as well as to simulate the pure rotational spectra.   

The pure rotational spectra are simulated using PGOPHER general purpose program \citep{WESTERN2017221}. At the second-order vibrational perturbation theory (VPT2) level, rotational-vibrational coupling, anharmonic and quartic centrifugal-distortion corrections were considered to obtain all the simulated spectra. PGOPHER is frequently used in astrophysical context on PANH molecules \citep{2020JQSRT.24907033N,2021A&A...652A.163C}. In PGOPHER, the Watson's A-reduced effective rotational Hamiltonian up to the sixth order terms were coupled with the nuclear quadrupole terms to obtain the transition frequencies and assigned rotational and hyperfine quantum numbers J$_{\rm K_{a}K_{c}}$ and F. Because of the existence of $^{14}$N(I = 1) nuclei, the lines in the cm region show resolved hyperfine splitting \citep{2018MNRAS.476.5268M}. Therefore, the $^{14}$N-nuclear quadrupole hyperfine structure is also simulated for the target molecules in the cm-wave region.
\section{Results and Discussion}
\subsection{\textbf{Accuracy of the computational methods}}
\subsubsection{Effects of different basis sets on rotational spectral parameters}
Beyond the rigid-rotor approximation, we briefly discuss the most frequent model Hamiltonians for determining the rotational spectroscopic constants of the chosen asymmetric top PANHs. The constants are first derived for the test molecules, benzonitrile, 1-cyanonaphthalene (1-CNN) and quinoline using three distinct basis sets-- B3LYP/6-311+G(d,p), B3LYP/cc-pVTZ and MP2/6-31G(d,p). Examining the rotational spectra of PANHs frequently involves the use of these basis sets \citep{2003JMoSp.217..115K,2018MNRAS.476.5268M}. 
The molecular constants calculated at the considered basis sets are given in Table 1 with the deviation between the experimental and theoretical rotational constants' values shown in parentheses. 

For all the test molecules, the ground state rotational constants (A$_{\rm 0}$, B$_{\rm 0}$ and C$_{\rm 0}$) computed at the B3LYP/6-311+G(d,p) level of theory agree within $\sim$0.6 per cent of those determined through experiments. These values are within $\sim$0.8 and $\sim$0.7 per cent of those obtained from the B3LYP/cc-pVTZ and MP2/6-31G(d,p) level of theory, respectively. The error values are the maximum percentage errors calculated on the rotational constants (Table 1). The mean percentage errors for rotational constants of the test molecules are 0.24, 0.63 and 0.28 with B3LYP/6-311+G(d,p), B3LYP/cc-pVTZ and MP2/6-31G(d,p), respectively. As a result, the ground state rotational constants of N- and CN-PAHs are better estimated by the B3LYP/6-311+G(d,p) level (Table 1).

All the level of computations used here show nearly equal performance in computing the centrifugal distortion constants ($\Delta_{\rm J}$, $\Delta_{\rm JK}$, $\Delta_{\rm K}$, $\delta_{\rm J}$ and $\delta_{\rm K}$). For CN containing PAHs, the computed values match within the mean percentage error of $\sim$6.7 per cent compared to the experiments by all the level of theories except $\Delta_{\rm JK}$ for 1-CNN. It shows relative error of $\sim$13.4 per cent (Table 1) with the B3LYP/6-311+G(d,p). The centrifugal distortion constants of N-PAH-quinoline are predicted to be within $\sim$3.4 per cent of the experimental values by all the level of theories considered here (Table 1). 

All the level of theories calculate the nuclear quadrupole coupling constants ($\chi_{xx}$, $\chi_{yy}$ and $\chi_{zz}$) to within $\sim$10 per cent of the experimental values, where the B3LYP/6-311+G(d,p) level achieves the best performance.
\begin{table}
\begin{threeparttable}
	\centering
	\caption{Rotational transitions of ground state of benzonitrile (C$_{6}$H$_{5}$CN), 1-cyanonaphthalene (C$_{10}$H$_{7}$CN) and quinoline (C$_{9}$H$_{7}$N) with relative error ($\sim\delta$).}
	\label{tab:example_table}
	\begin{tabular}{lcccc} 
		\hline
		\hline
		\multicolumn{2}{c}{Transition} &
		\multicolumn{2}{c}{Frequency} & $\delta$\\
		\cmidrule(lr){1-2}
		\cmidrule(lr){3-4}

		J$'_{K'_a,K'_c}$ $\rightarrow$ J"$_{K"_a,K"_c}$ & $F' \rightarrow F"$ & Experimental$^{\rm \dagger}$ & Calculated$^{\rm \ddagger}$ & \\ 
		 & & (MHz) & (MHz) & (\%)\\
		\hline
	\hspace{-0.25cm} C$_{6}$H$_{5}$CN &&&&\\	
	$7_{0,7} \rightarrow 6_{0,6}$	& 6 $\rightarrow$ 5 & 18409.3490 & 18438.2834 & 0.15\\
	$7_{2,6} \rightarrow 6_{2,5}$ & 7 $\rightarrow$ 6 & 19178.9991 & 19203.9883 & 0.13\\
	$7_{2,5} \rightarrow 6_{2,4}$	& 8 $\rightarrow$ 7 & 20096.0931 & 20115.8509 & 0.09\\
	$7_{1,6} \rightarrow 6_{1,5}$	& 8 $\rightarrow$ 7 & 20192.9641 & 20217.2728 & 0.12\\
	$8_{0,8} \rightarrow 8_{0,7}$	& 8 $\rightarrow$ 7 & 20828.2065 & 20861.7296 & 0.16\\
	$8_{2,7} \rightarrow 7_{2,6}$	& 8 $\rightarrow$ 7 & 21855.9330 & 21884.8986 & 0.13\\
	$8_{1,7} \rightarrow 7_{1,6}$	& 8 $\rightarrow$ 7 & 22943.4750 & 22972.4289 & 0.12\\
	$9_{0,9} \rightarrow 8_{0,8}$	& 8 $\rightarrow$ 7 & 23227.6869 & 23265.5033 & 0.16\\
	\hspace{-0.25cm} C$_{10}$H$_{7}$CN &&&&\\
	$9_{0,9} \rightarrow 8_{0,8}$	& 9 $\rightarrow$ 8   & 11039.2815 & 11020.1694 & 0.17\\
	$9_{2,8} \rightarrow 8_{2,7}$	& 9 $\rightarrow$ 8  & 12202.9932 & 12181.7064 & 0.17\\
	$9_{2,7} \rightarrow 8_{2,6}$	& 10 $\rightarrow$ 9  & 13481.0643 & 13471.1245 & 0.07\\
	$10_{3,8} \rightarrow 9_{3,7}$	& 10 $\rightarrow$ 9  & 14525.9246 & 14498.4194 & 0.18\\
	\hspace{-0.25cm} C$_{9}$H$_{7}$N &&&&\\
	$4_{2,3} \rightarrow 3_{1,2}$	  & 4 $\rightarrow$ 3 & 13777.0811 & 13807.7524 & 0.22\\	
	$14_{3,11} \rightarrow 14_{2,12}$	  & 14 $\rightarrow$ 14 & 14510.9656 & 14469.6863 & 0.28\\
	$18_{4,14} \rightarrow 18_{3,15}$	  & 19 $\rightarrow$ 19 & 17853.9280 & 17786.8871 & 0.37\\
	$28_{7,21} \rightarrow 28_{6,22}$	  & 22 $\rightarrow$ 22 & 22156.2876 & 22030.2369 & 0.57\\
	$18_{7,12} \rightarrow 18_{6,13}$	  & 18 $\rightarrow$ 18 & 25309.5059 & 25433.3977 & 0.48\\
	$13_{2,12} \rightarrow 12_{1,11}$	  & 12 $\rightarrow$ 11 & 26422.1001 & 26439.6661 & 0.06\\
	\hline
    \end{tabular}
    \begin{tablenotes}
    \item $^{\rm \dagger}$The experimental data are taken from \cite{2018Sci...359..202M} for C$_{6}$H$_{5}$CN; \cite{2018MNRAS.476.5268M} for C$_{10}$H$_{7}$CN and \cite{2003JMoSp.217..115K} for C$_{9}$H$_{7}$N. 
    \item $^{\rm \ddagger}$B3LYP/6-311+G(d,p).
    \end{tablenotes}
    \end{threeparttable}
    \end{table}
    
    \begin{figure}
        \centering
        \includegraphics[width=\columnwidth]{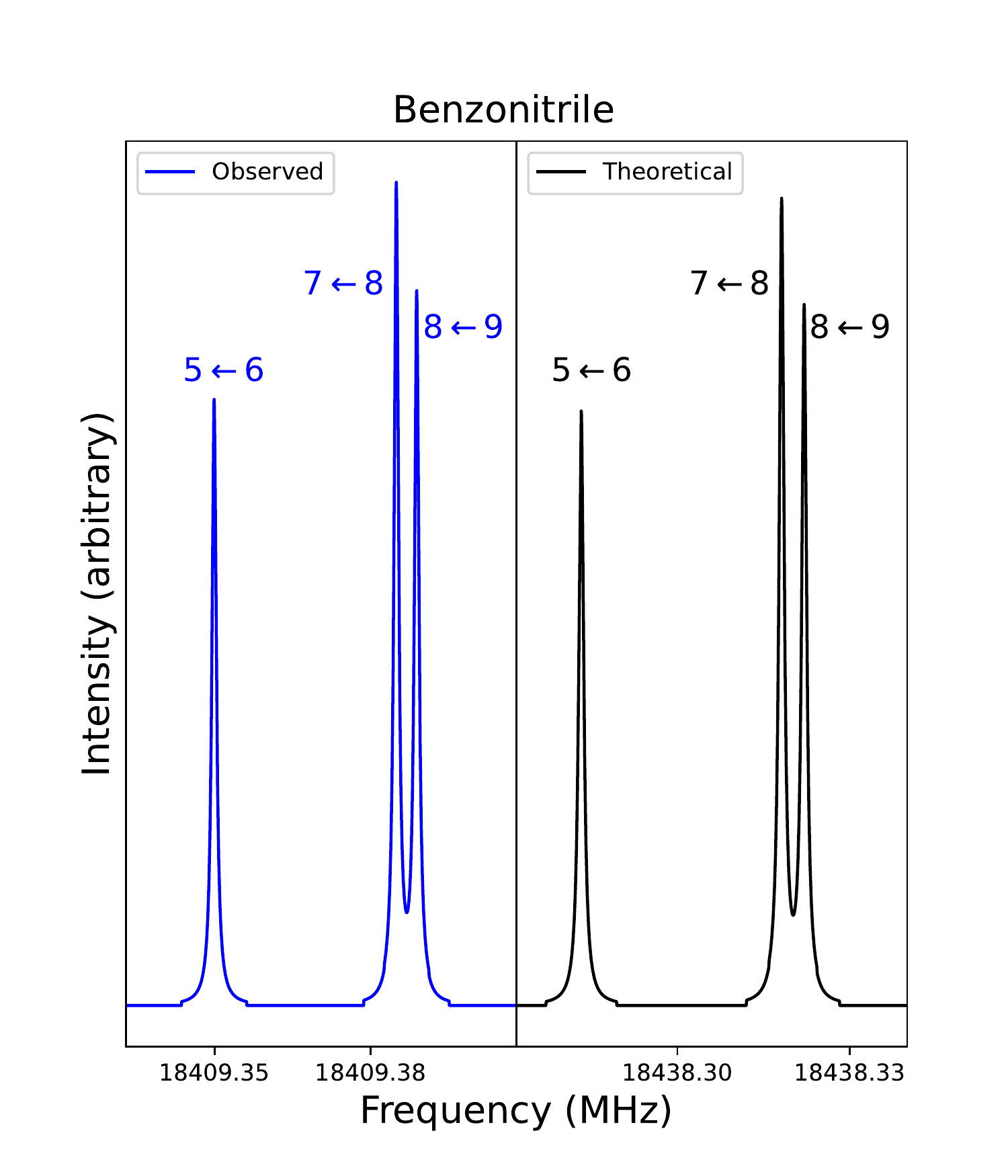}
        \caption{14$^{\rm N}$-nuclear quadrupole hyperfine components of the $7_{0,7} \rightarrow 6_{0,6}$ transition with $F" \rightarrow F'$ shown in blue and black colors for the observed and simulated benzonitrile transition, respectively, with the B3LYP/6-311+G(d,p) level.}
        \label{fig:my_label}
    \end{figure}
  \begin{figure*}
     \centering
     \begin{subfigure}[b]{0.48\textwidth}
         \centering
         \includegraphics[width=\textwidth]{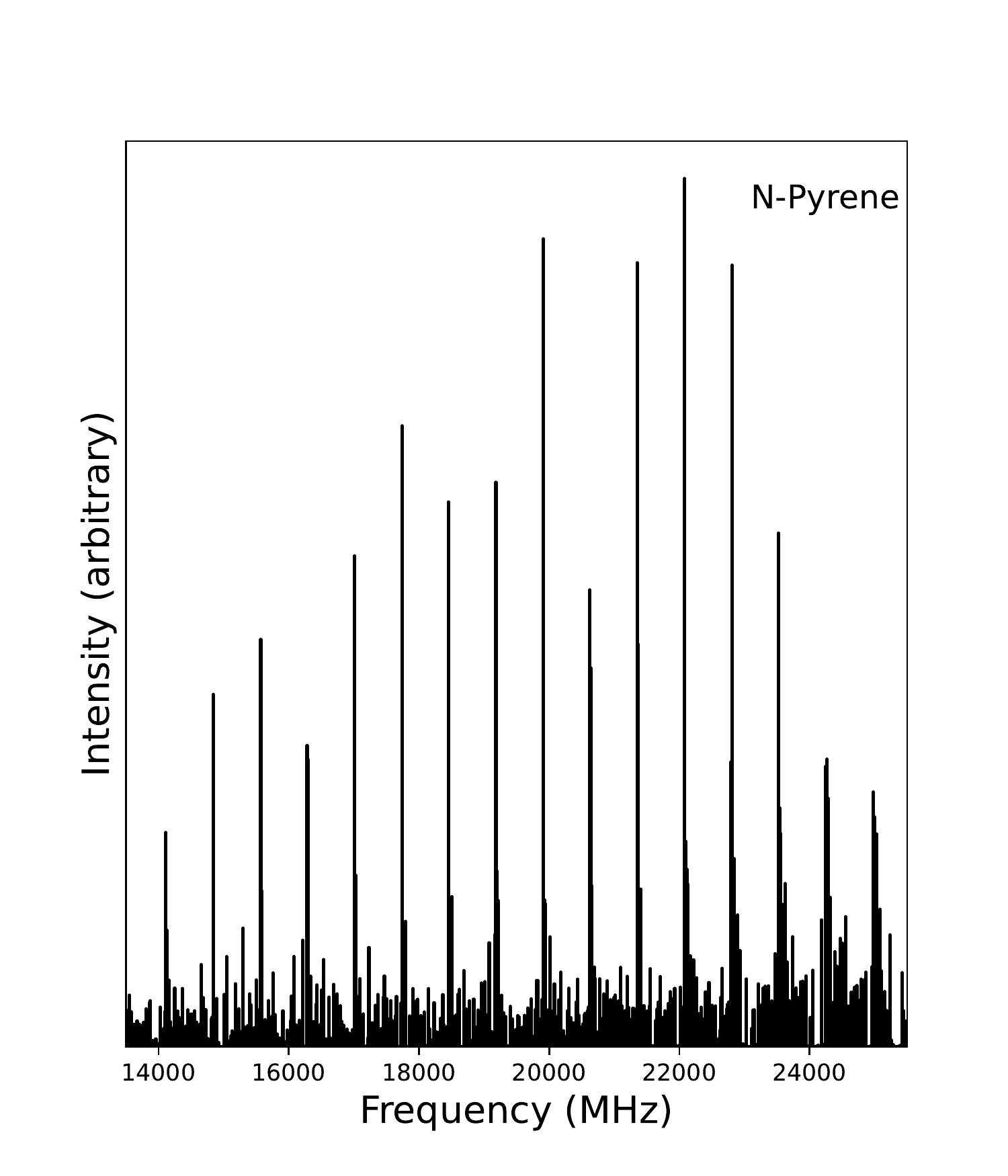}
         \caption{(a)}
     \end{subfigure}
     \begin{subfigure}[b]{0.48\textwidth}
         \centering
         \includegraphics[width=\textwidth]{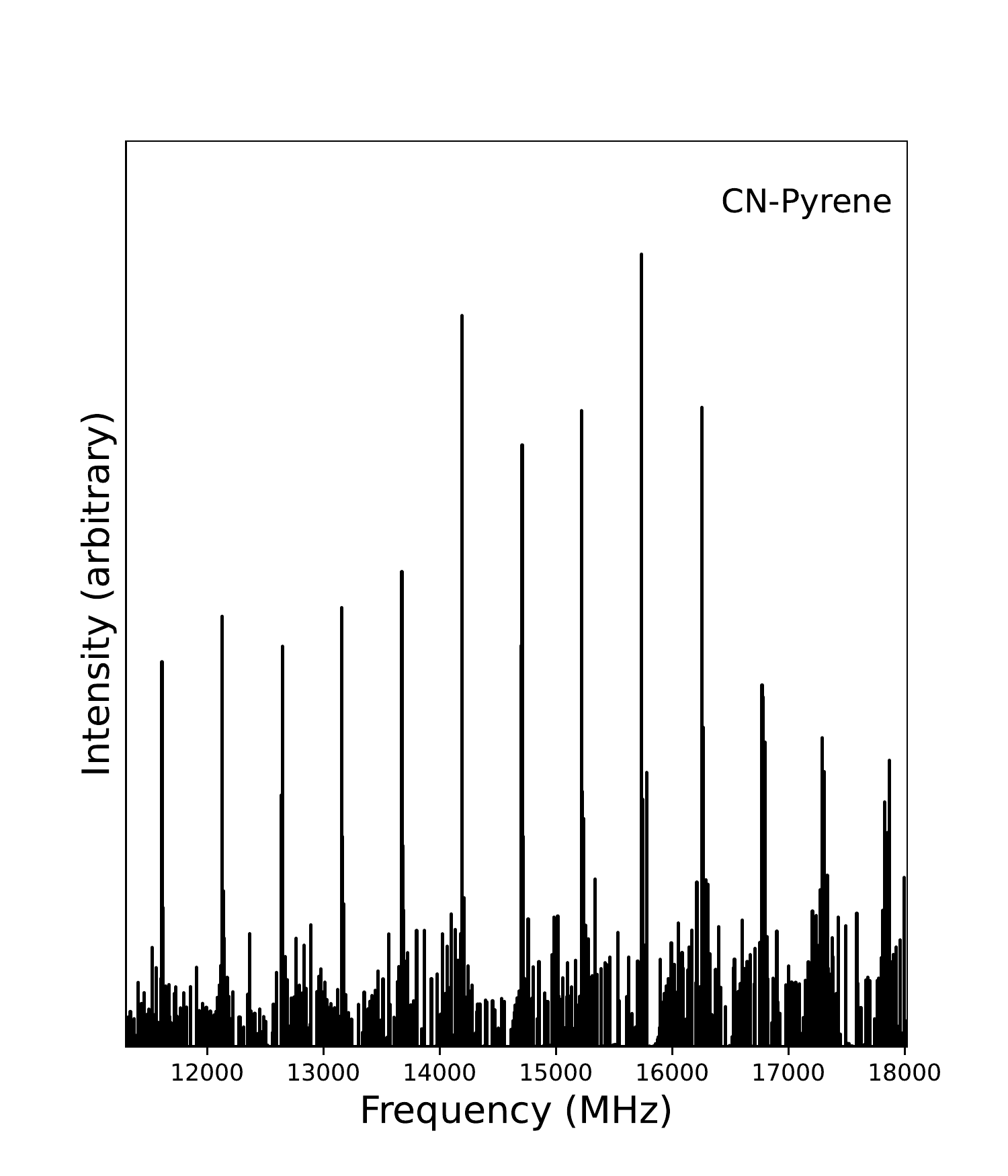}
         \caption{(b)}
         \label{(b)}
     \end{subfigure}
          \caption{Rotational spectrum simulation of pyrene in N-PAH and CN-PAH variants at 300 K with B3LYP/6-311+G(d,p) level.}
\end{figure*}
\begin{figure*}
     \centering
         \begin{subfigure}[b]{0.48\textwidth}
         \centering
         \includegraphics[width=\textwidth]{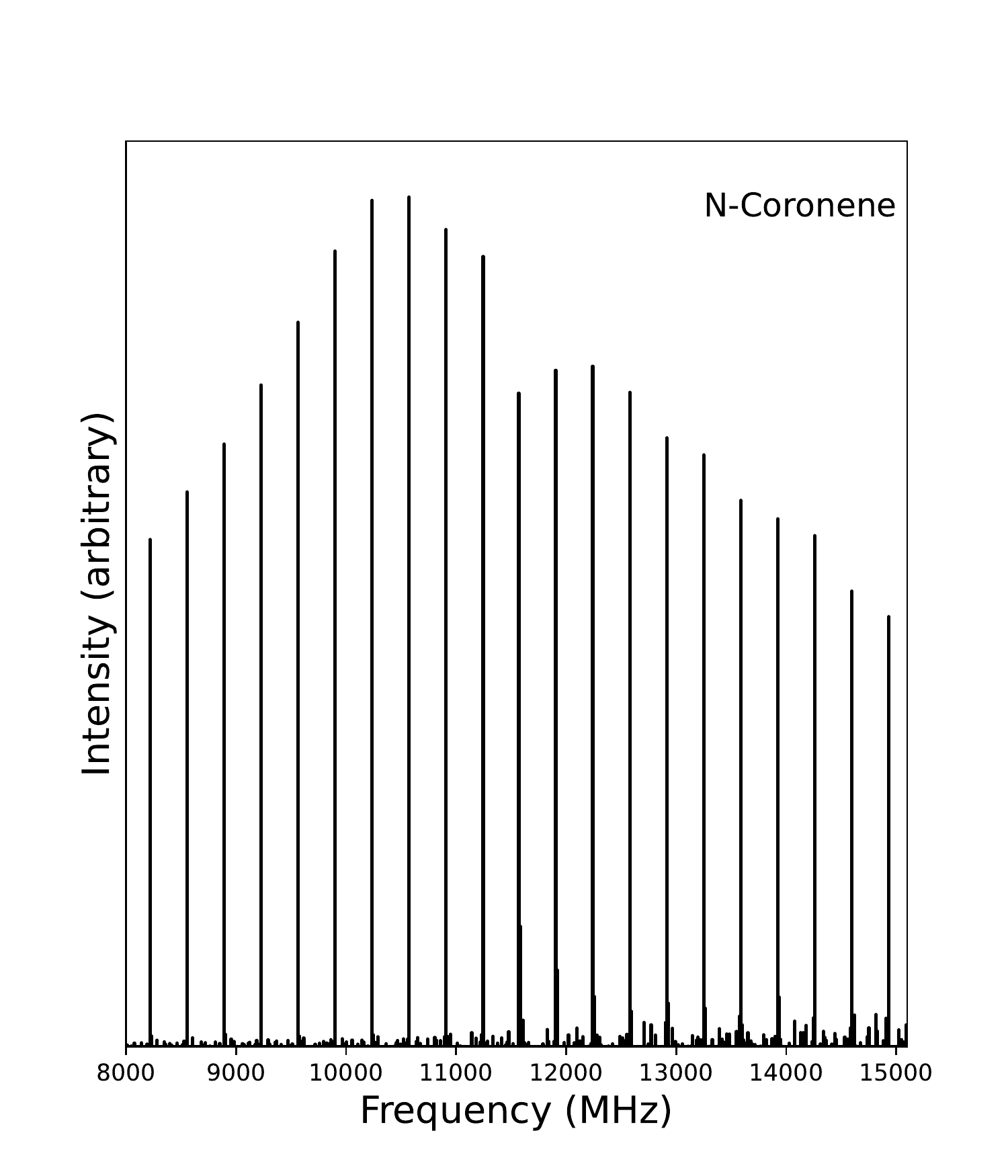}
         \caption{(a)}
         \label{fig:three sin x}
     \end{subfigure}
         \vspace{1cm}
         \begin{subfigure}[b]{0.48\textwidth}
         \centering
         \includegraphics[width=\textwidth]{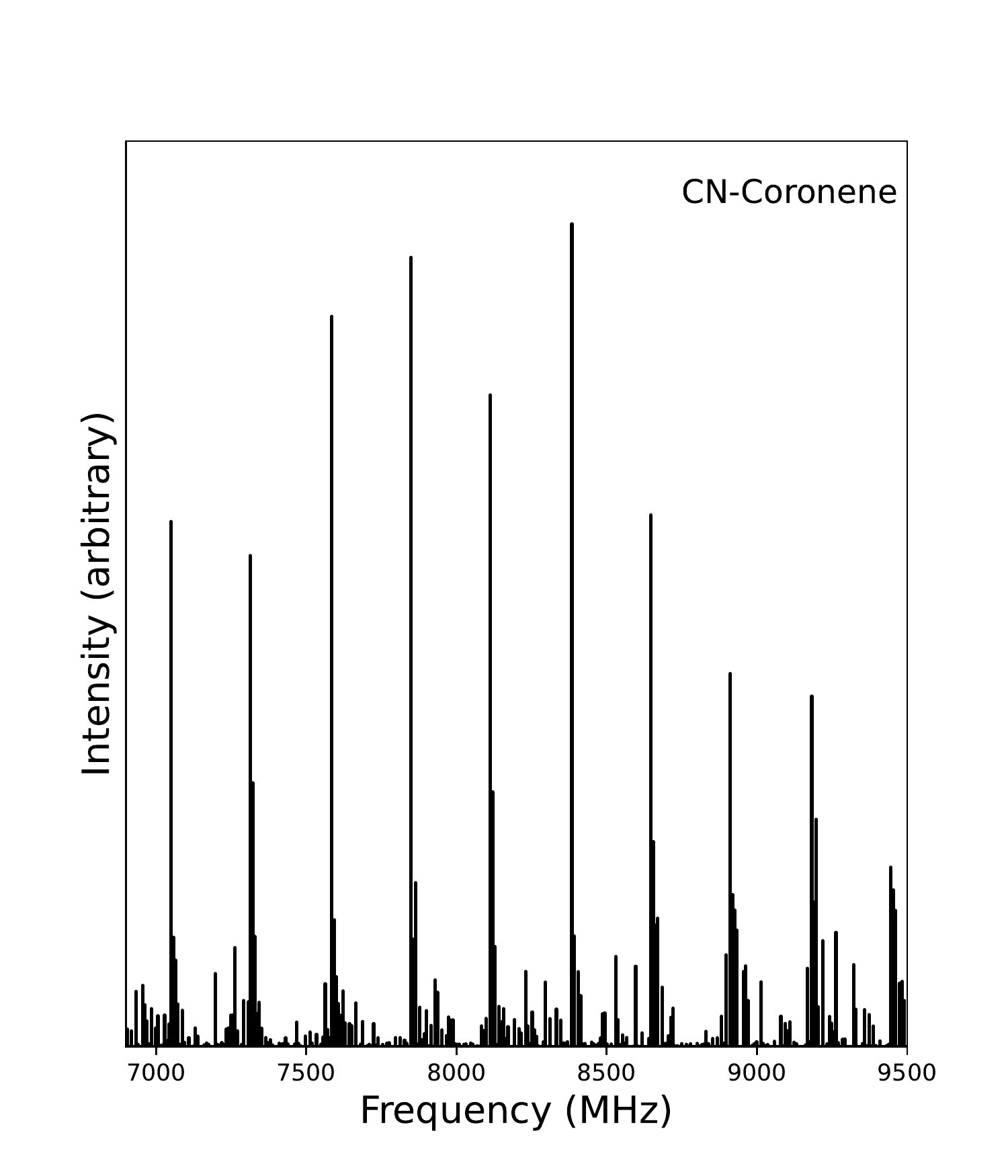}
         \caption{(b)}
         \label{fig:three sin x}
     \end{subfigure}     
     \caption{Same as Fig. 4 for Coronene.}
\end{figure*}  
\subsubsection{Simulation of rotational spectra with hyperfine splitting}
The pure rotational spectra of benzonitrile, 1-CNN and quinoline are shown in Figure 2. These are simulated at 300 K with the B3LYP/6-311+G(d,p) level of theory, in comparison with the experimental spectra in the cm-wave region. Some of the rotational transitions of the test molecules are tabulated in Table 2 with relative error in the experimental and calculated values at the B3LYP/6-311+G(d,p) level. Besides this, the $^{14}$N-nuclear quadrupole hyperfine structure for an observed benzonitrile transition, $7_{0,7} \rightarrow 6_{0,6}$ (J$'_{K'_a,K'_c}$ $\rightarrow$ J"$_{K"_a,K"_c}$ with F = J + I) in the TMC-1 is also shown in Figure 3.  

In comparison with the experimental values, the transition frequencies are computed to within $\sim$0.18 per cent for the CN-PAH and within $\sim$0.5 per cent for the N-PAH considered here (Table 2 \& Figure 2). Even small variations in the rotational constants' precision in spectral simulations have a big impact on how well the line frequencies are predicted. The perusal of Table 2 reveals that the discrepancies between the experimental and theoretical transitions are larger for the N-PAH (C$_{9}$H$_{7}$N) despite having more accurate prediction of rotational constants (A$_0$, B$_0$, C$_0$) compared to the CN-PAHs (Table 1). This indicates a significant dominance of the A$_0$ rotational constant in the simulation of the N-PAH rotational spectrum, which, in the case of CN-PAHs, appears to be the least dominating. 

The $^{14}$N-nuclear quadrupole hyperfine structure simulation of $7_{0,7} \rightarrow 6_{0,6}$ benzonitrile observed transition shown in Figure 3 is well resolved, which proves the usefulness of the applied method in simulating the hyperfine structure of the PANH molecules that have been taken into consideration. 

Based on the above, it is evident that the B3LYP/6-311+G(d,p) level of theory is an appropriate level for estimating molecular constants and simulating rotational spectrum with hyperfine structure for both N-PAH and CN-PAH variants. As a result, in the following section, two comparatively large PAHs, pyrene (C$_{16}$H$_{10}$) and coronene (C$_{24}$H$_{12}$) for both the N and CN substituted variants are investigated with the B3LYP/6-311+G(d,p) level of theory. 
\subsection{Pure rotational spectroscopy of target PAHs}
N-PAHs are believed to represent a substantial portion of the astronomical PAH population and CN-PAHs have already been detected in the ISM. They both exhibit strong polarity and stand out as the most potential target for pure rotational spectroscopy. In this light, based on the finding examined in the previous section, the molecular constants and rotational spectra of pyrene (C$_{16}$H$_{10}$) and coronene (C$_{24}$H$_{12}$) in both the N-PAH and CN-PAH variants are given with the B3LYP/6-311+G(d,p) level of theory. It is apparent that the dipole moment components for each of these planar molecules are restricted to the plane of the PAH, i.e., $\mu_c$=0 (Table 3). 
\begin{table*}\normalsize
\begin{threeparttable}
\centering
        \caption{Rotational spectral parameters of pyrene and coronene in N-PAH and CN-PAH variant with B3LYP/6-311+G(d,p) level.}
    \begin{tabular}{lcccc}
    \hline
    \hline
   Constants & \multicolumn{2}{c}{Pyrene} & \multicolumn{2}{c}{Coronene}\\
   \cmidrule(lr){2-3}
   \cmidrule(lr){4-5}
  & N-PAH & CN-PAH & N-PAH & CN-PAH\\\hline
 A$_{\rm 0}^{\rm a}$ (MHz) & 1018.8042 & 846.1307 & 337.4456 & 333.9216\\     
 B$_{\rm 0}$ (MHz) & 561.3018  & 371.0981 & 333.6118 & 221.1045\\
 C$_{\rm 0}$ (MHz) & 361.9103 & 257.9608 & 167.7588 & 133.0236\\
 $\Delta_{\rm J}^{\rm b}$ (Hz) & 2.25 & 1.91 & 0.89 & 0.20\\ 
 $\Delta_{\rm JK}$ (Hz) & 1.09 & -4.9 & -1.44 & 0.24\\ 
 $\Delta_{\rm K}$ (Hz) & 11.51 & 19 & 0.62 & 0.81\\ 
 $\delta_{\rm J}$ (Hz) & 0.78 & 0.69 & 0.008 & 0.071\\ 
 $\delta_{\rm K}$ (Hz) & 4.07 & 3.11 & 0.53 & 0.61\\
 $\chi_{xx}^{\rm c}$ (MHz) & -4.412 & 2.365 & -4.844 & 2.407\\
 $\chi_{yy}$ (MHz) & 3.362 & 2.034 & 3.383 & 1.935\\
 $\chi_{zz}$ (MHz) & 1.428 & -4.399 & 1.461 & -4.342\\
  $\mu^{\rm d}$ (Debye) & $\mu_{\rm a}$ = 1.8904 & $\mu_{\rm a}$ = 3.6716 & $\mu_{\rm a}$ = 2.2535 & $\mu_{\rm a}$ = 1.1782\\
     & $\mu_{\rm b}$ = 2.1205 & $\mu_{\rm b}$ = 4.0694 & $\mu_{\rm b}$ = 1.5565 & $\mu_{\rm b}$ = 5.4072\\
 & $\mu_{\rm tot.}$ = 2.8408 & $\mu_{\rm tot.}$ = 5.4809 & $\mu_{\rm tot.}$ = 2.7388 & $\mu_{\rm tot.}$ = 5.5341\\
\hline
 \end{tabular}
 \label{tab:my_label}
 	\begin{tablenotes}
	\item Notes. $^{\rm a}$Rotational constants, $^{\rm b}$Quartic centrifugal distortion constants, $^{\rm c}$$^{14}$N quadrupole coupling constants in reference to the principal  axis system, $^{\rm d}$Dipole moment.
\end{tablenotes}
\end{threeparttable}
\end{table*}
\subsubsection{Pyrene}
The molecular constants of pyrene in N-PAH and CN-PAH variants are given in Table 3 and the simulation of a portion of the cm-wave spectra is shown in Figure 4. In pyrene, the ground state rotational constants (A$_0$, B$_0$, C$_0$) are large for N-variant compared to the CN-variant. For example, the A$_0$ rotational constant is 1018.8042 MHz for N-pyrene while for CN-pyrene, the value is 846.1307 MHz (Table 3). The average difference between the rotational constants of N- and CN-pyrene is $\sim$150 MHz, where the maximum difference is observed for the B$_0$. With the exception of the $\Delta_{\rm K}$, the centrifugal distortion constants are also large for N-pyrene. The $^{14}$N quadrupole coupling constants of N-pyrene and CN-pyrene show similar values to their counterpart species given in Table 1 (quinoline and 1-CNN, respectively). For example, the $\chi_{zz}$ is 1.428 MHz for N-pyrene and 1.462 MHz for quinoline, for CN-pyrene, the $\chi_{zz}$ differs with 1-CNN by only 0.07 MHz. These small changes in $\chi$(N) clearly indicates that the N atom's local electronic structure is independent from the proximity of the atomic ring. CN-PAHs seem to be strongly polar compared to similar-sized N-PAHs. The magnitude of the dipole moments are larger for CN-pyrene, which is consistent with the trend shown in Table 1 for the N-PAH (quinoline) and CN-PAH (1-CNN). 

The region with the strongest transitions at 300 K in the cm-wave spectra of N-pyrene and CN-pyrene are shown in Figure 4(a) and Figure 4(b), respectively. The strongest lines fall in the K (18--27 GHz) band for N-pyrene, whereas for CN-pyrene, the strongest transitions are found in the K$_{\rm u}$ (12--18 Ghz) band. At 300 K, the strongest line appears around 22 GHz for N-pyrene (Figure 4-a) and near 16 GHz for CN-pyrene (Figure 4-b). 
\subsubsection{Coronene}
Table 3 presents the molecular constants of N-coronene and CN-coronene while their rotational spectra in the cm-wave region are presented in Figure 5. The rotational constants are large for N-PAH variant for coronene as well, however, the average difference in the rotational constants between N- and CN-coronene is $\sim$50 MHz, which is lower than the difference seen for pyrene (Table 3). Similar to pyrene, the rotational constant B$_0$ in coronene exhibits the maximum difference between its N- and CN-PAH variant. In coronene, the centrifugal distortion constants are higher for the CN-PAH variant except the $\Delta_{\rm J}$. The $^{14}$N quadrupole coupling constants of N-coronene and CN-coronene show similar behavior to pyrene (Table 3). The CN-PAH version is more polar compared to the N-PAH for coronene as well.

The cm-wave spectra of coronene with strongest lines at 300 K are displayed in Figure 5(a) for N-PAH while in Figure 5(b) for CN-PAH. For both the N- and CN-coronene, the strongest transitions lie within X (8--12 GHz) band. The strongest lines at 300 K are found near 11 GHz and 8.5 GHz for N-coronene (Figure 5-a) and CN-coronene (Figure 5-b), respectively.  
\begin{figure*}
     \centering
         \begin{subfigure}[b]{0.331\textwidth}
         \centering
         \includegraphics[width=\textwidth]{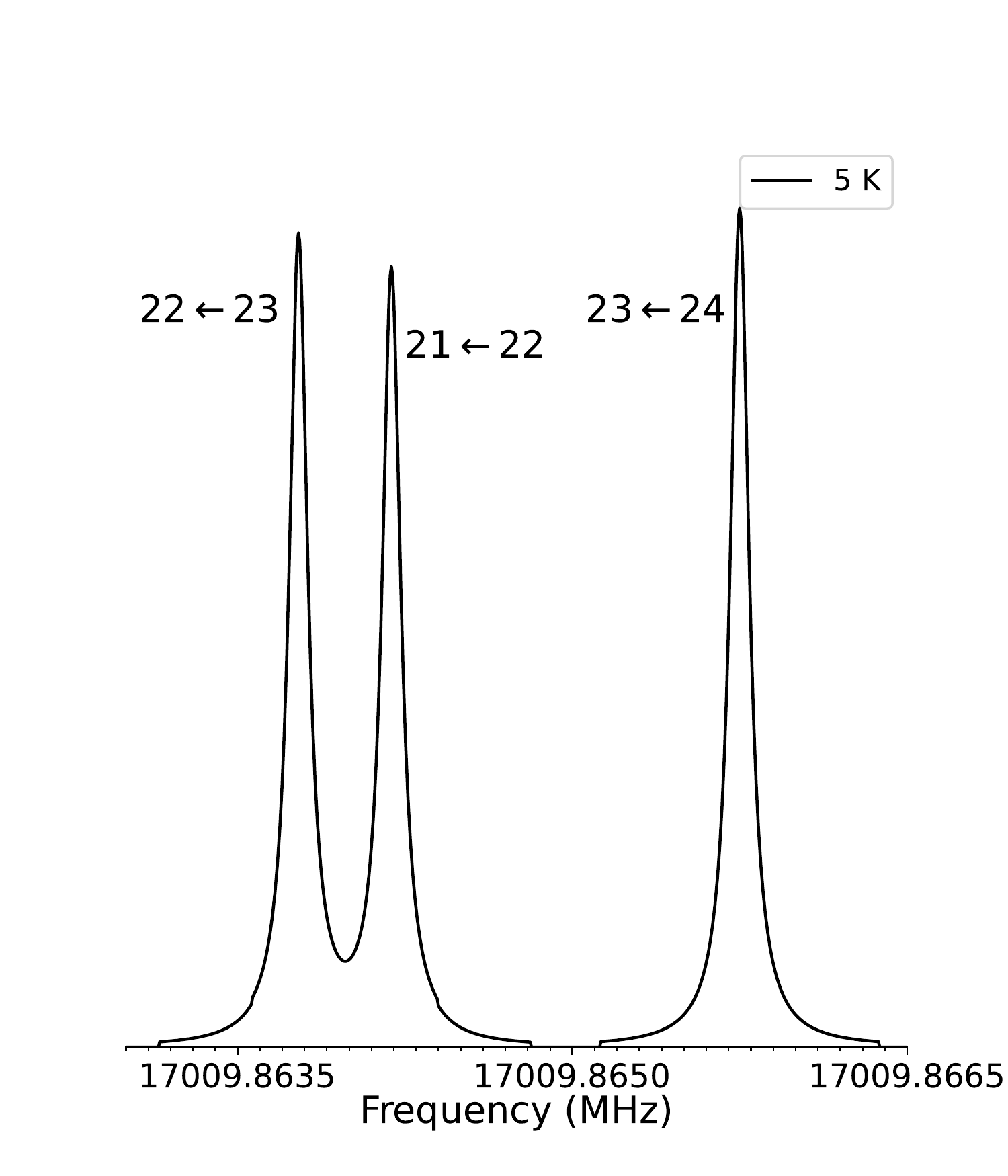}
         \caption{J$'_{K'_a,K'_c}$ $\rightarrow$ J"$_{K"_a,K"_c}$ = 23$_{1,23}$ $\rightarrow$ 22$_{0,22}$}
         \label{}
     \end{subfigure}
         \begin{subfigure}[b]{0.331\textwidth}
         \centering
         \includegraphics[width=\textwidth]{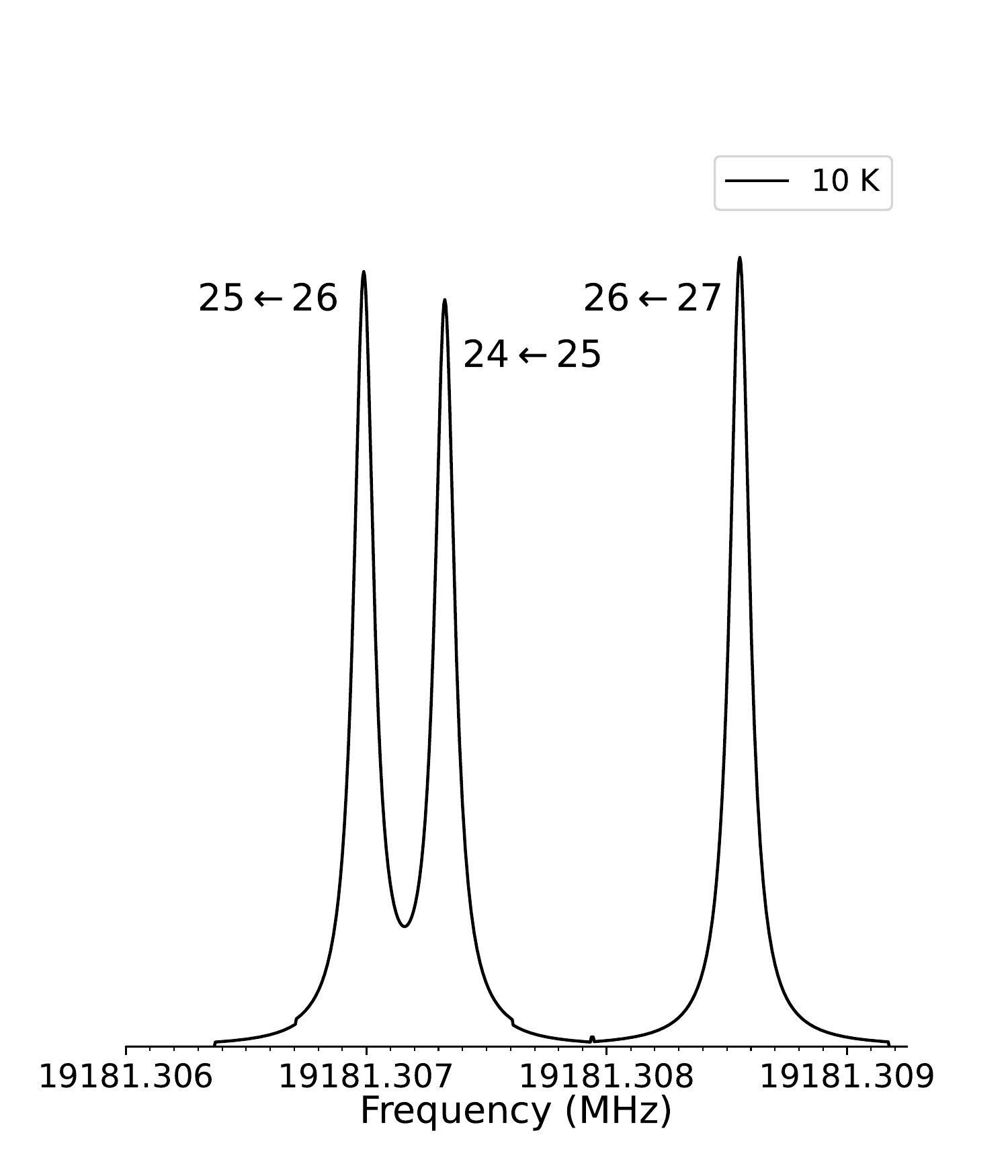}
         \caption{J$'_{K'_a,K'_c}$ $\rightarrow$ J"$_{K"_a,K"_c}$ = 26$_{1,26}$ $\rightarrow$ 25$_{0,25}$}
         \label{}
     \end{subfigure}  
         \begin{subfigure}[b]{0.331\textwidth}
         \centering
         \includegraphics[width=\textwidth]{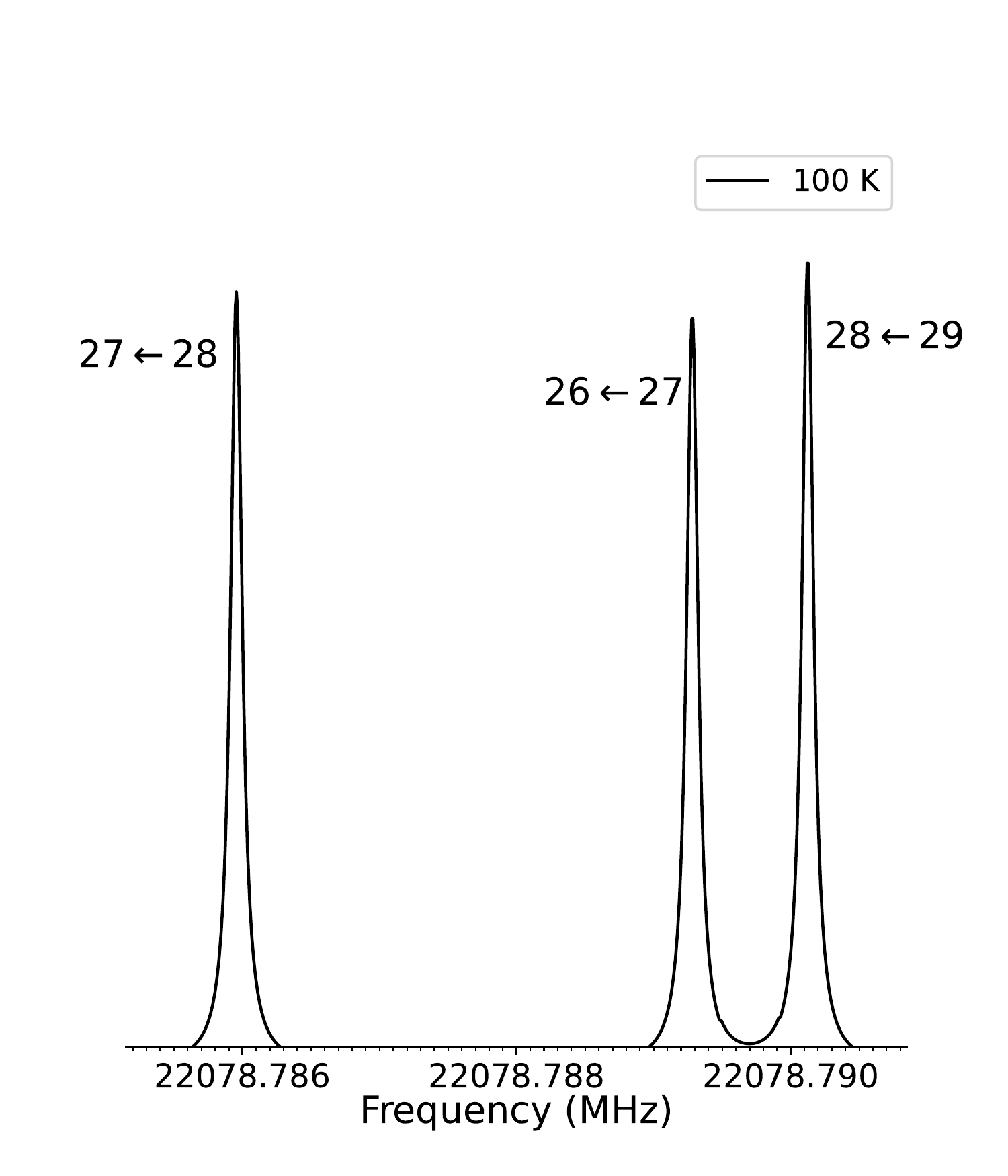}
         \caption{J$'_{K'_a,K'_c}$ $\rightarrow$ J"$_{K"_a,K"_c}$ = 28$_{3,26}$ $\rightarrow$ 27$_{2,25}$}
         \label{}
     \end{subfigure}  
     \caption{The strongest transition of N-pyrene showing resolved hyperfine structure at 5, 10 and 100 K temperatures without accounting for Doppler splitting.}
\end{figure*}
\begin{figure*}
     \centering
         \begin{subfigure}[b]{0.331\textwidth}
         \centering
         \includegraphics[width=\textwidth]{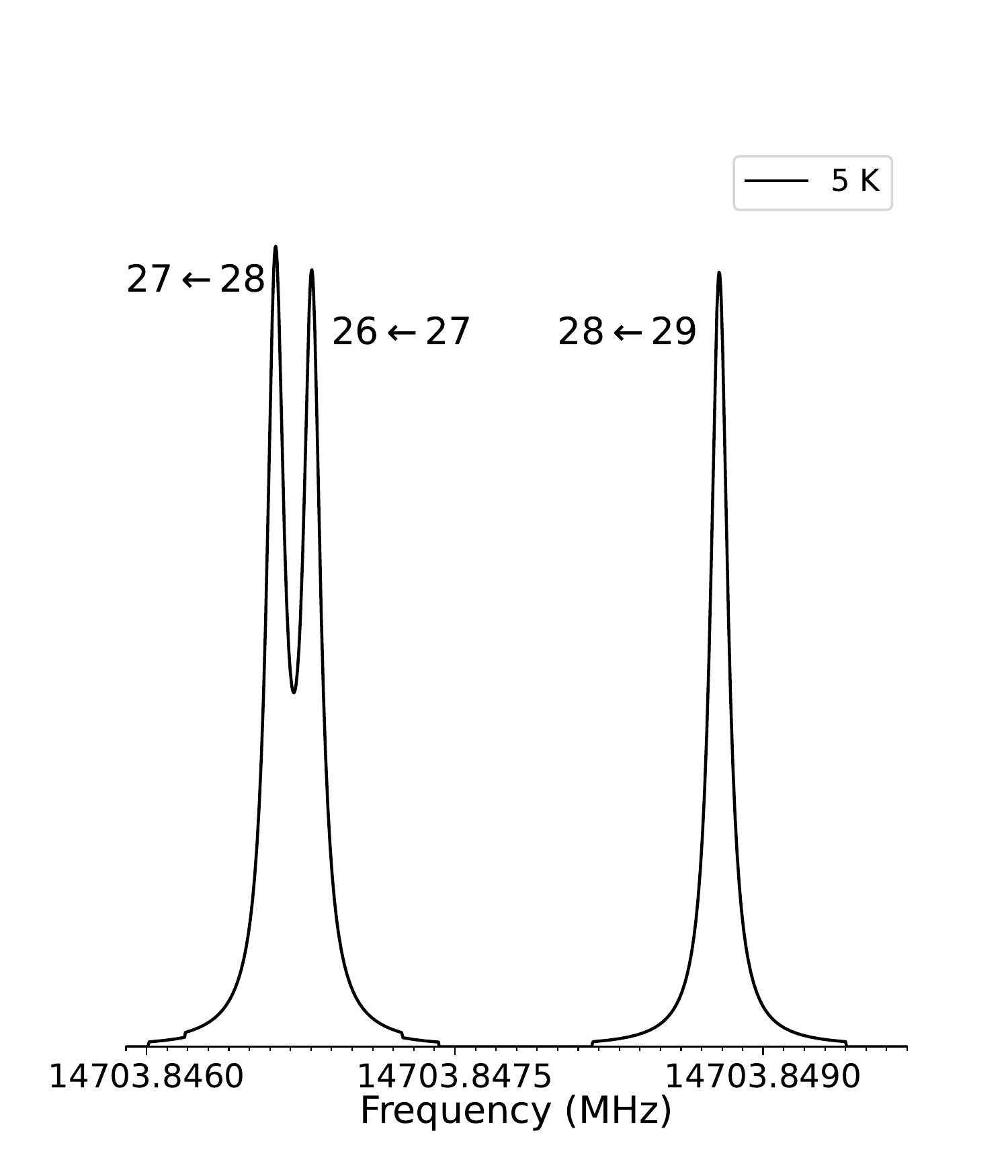}
         \caption{J$'_{K'_a,K'_c}$ $\rightarrow$ J"$_{K"_a,K"_c}$ = 28$_{1,28}$ $\rightarrow$ 27$_{0,27}$}
         \label{}
     \end{subfigure}
         \begin{subfigure}[b]{0.331\textwidth}
         \centering
         \includegraphics[width=\textwidth]{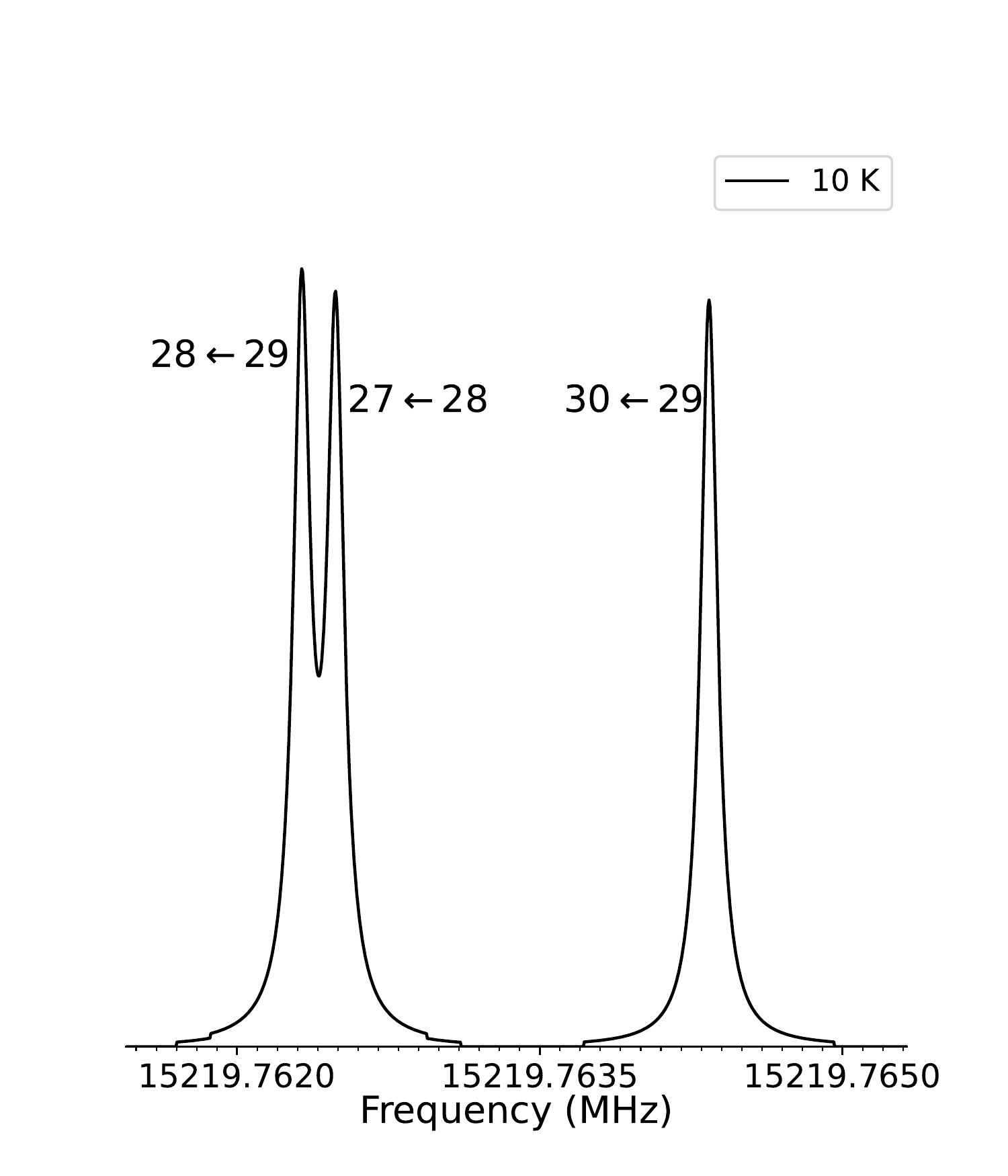}
         \caption{J$'_{K'_a,K'_c}$ $\rightarrow$ J"$_{K"_a,K"_c}$ = 29$_{1,29}$ $\rightarrow$ 28$_{0,28}$}
         \label{}
     \end{subfigure}  
         \begin{subfigure}[b]{0.331\textwidth}
         \centering
         \includegraphics[width=\textwidth]{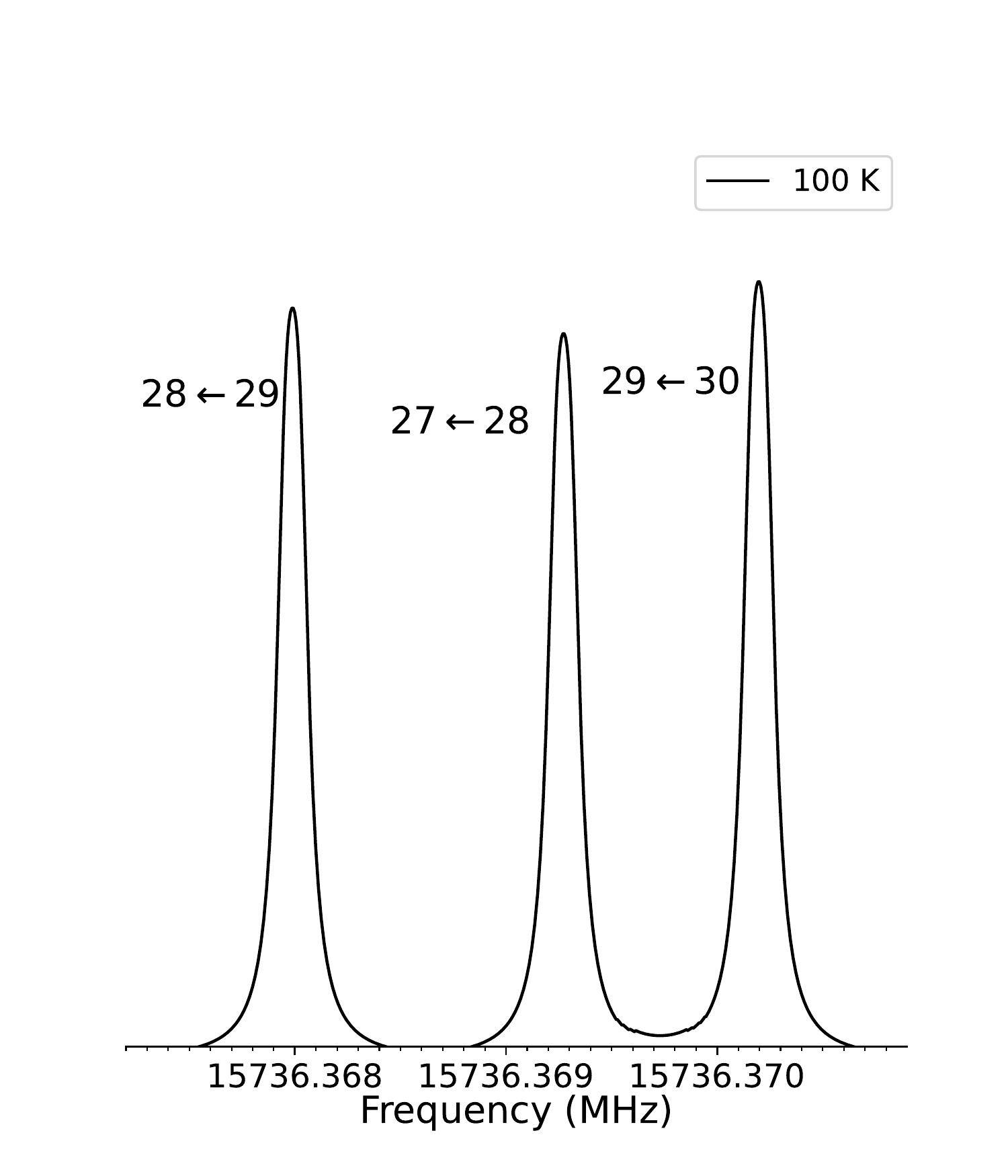}
         \caption{J$'_{K'_a,K'_c}$ $\rightarrow$ J"$_{K"_a,K"_c}$ = 29$_{2,28}$ $\rightarrow$ 28$_{1,27}$}
         \label{}
     \end{subfigure}  
     \caption{Same as Figure 6 for CN-pyrene.}
\end{figure*}
\begin{figure*}
     \centering
         \begin{subfigure}[b]{0.331\textwidth}
         \centering
         \includegraphics[width=\textwidth]{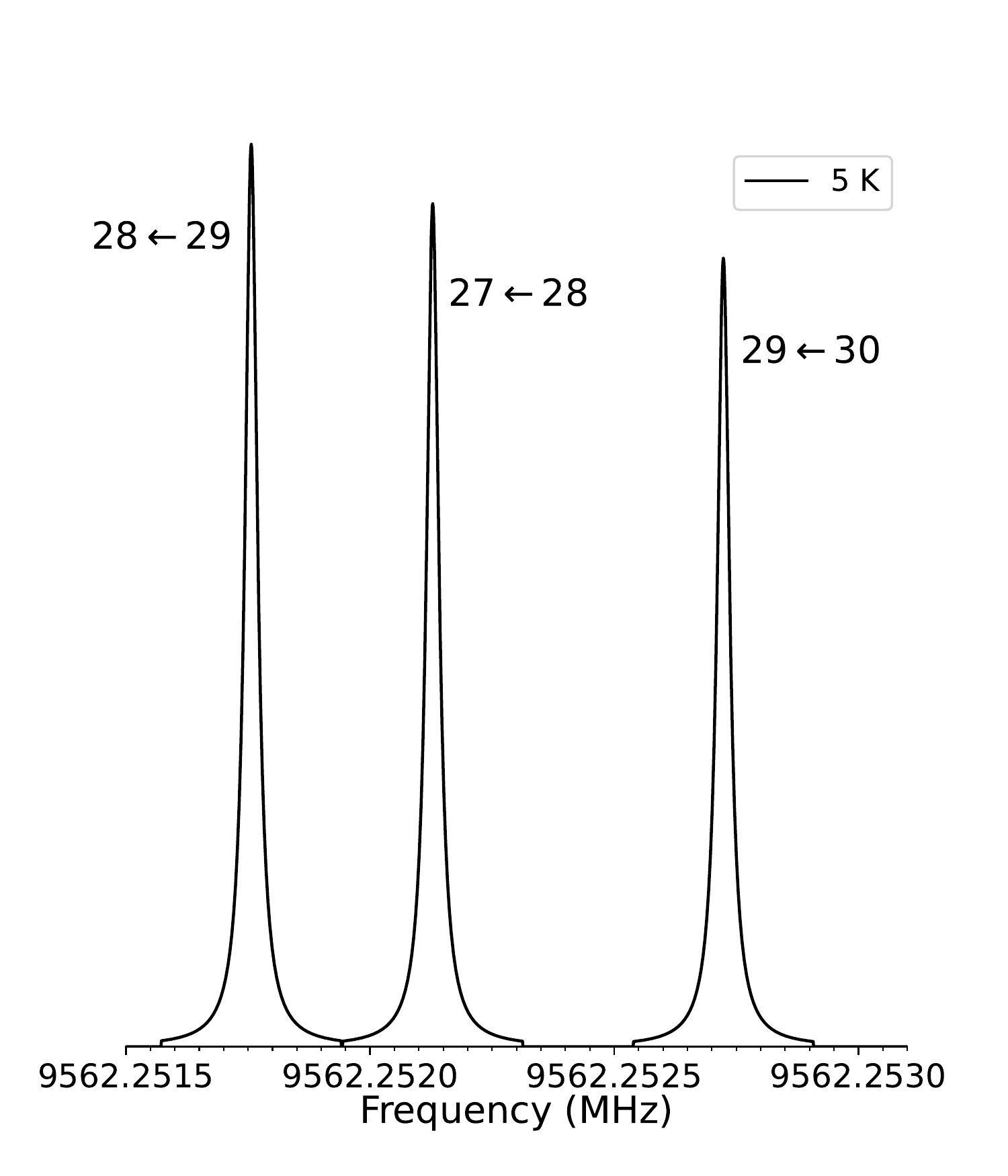}
         \caption{J$'_{K'_a,K'_c}$ $\rightarrow$ J"$_{K"_a,K"_c}$ = 28$_{1,28}$ $\rightarrow$ 27$_{1,27}$}
         \label{}
     \end{subfigure}
         \begin{subfigure}[b]{0.331\textwidth}
         \centering
         \includegraphics[width=\textwidth]{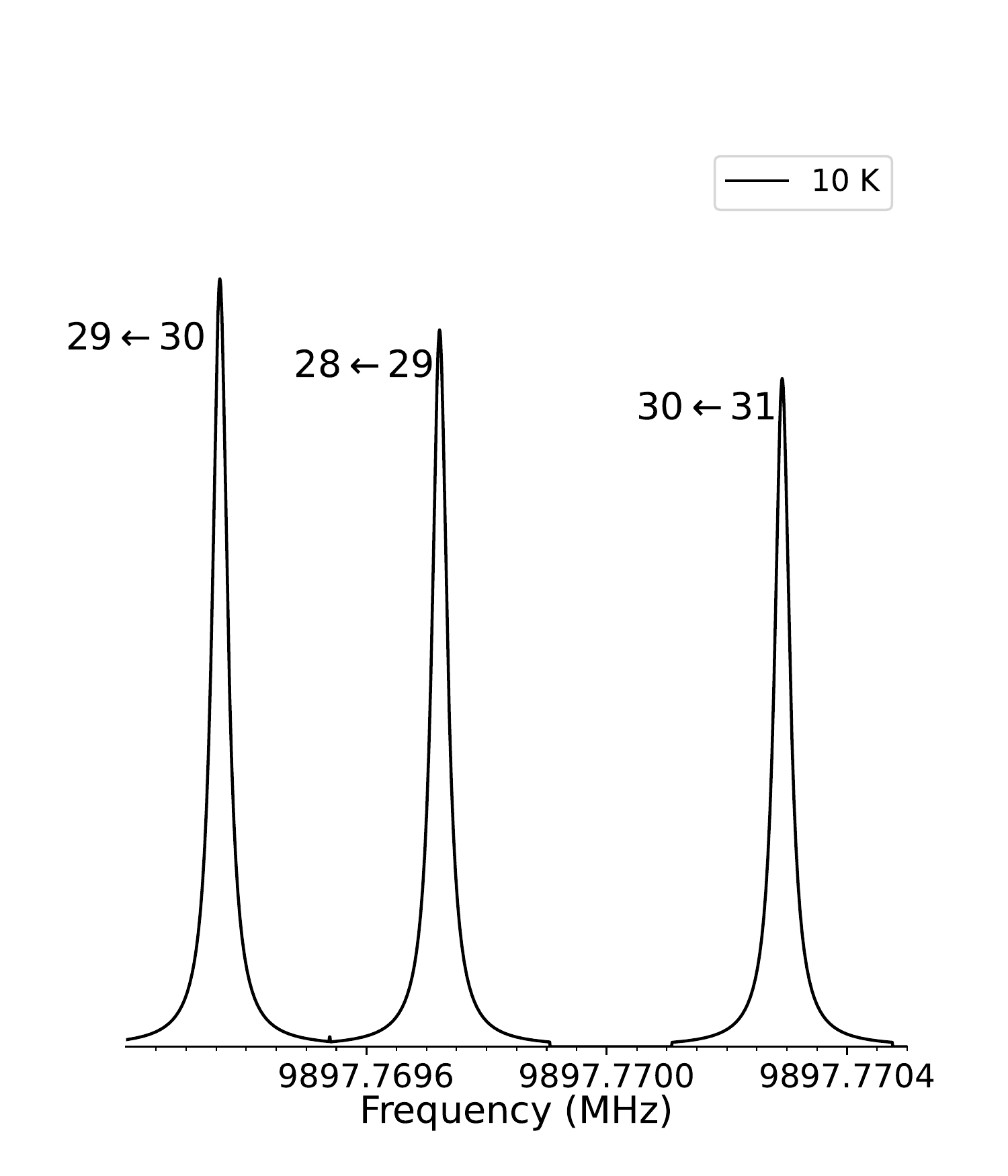}
         \caption{J$'_{K'_a,K'_c}$ $\rightarrow$ J"$_{K"_a,K"_c}$ = 29$_{0,29}$ $\rightarrow$ 28$_{0,28}$}
         \label{}
     \end{subfigure}  
         \begin{subfigure}[b]{0.331\textwidth}
         \centering
         \includegraphics[width=\textwidth]{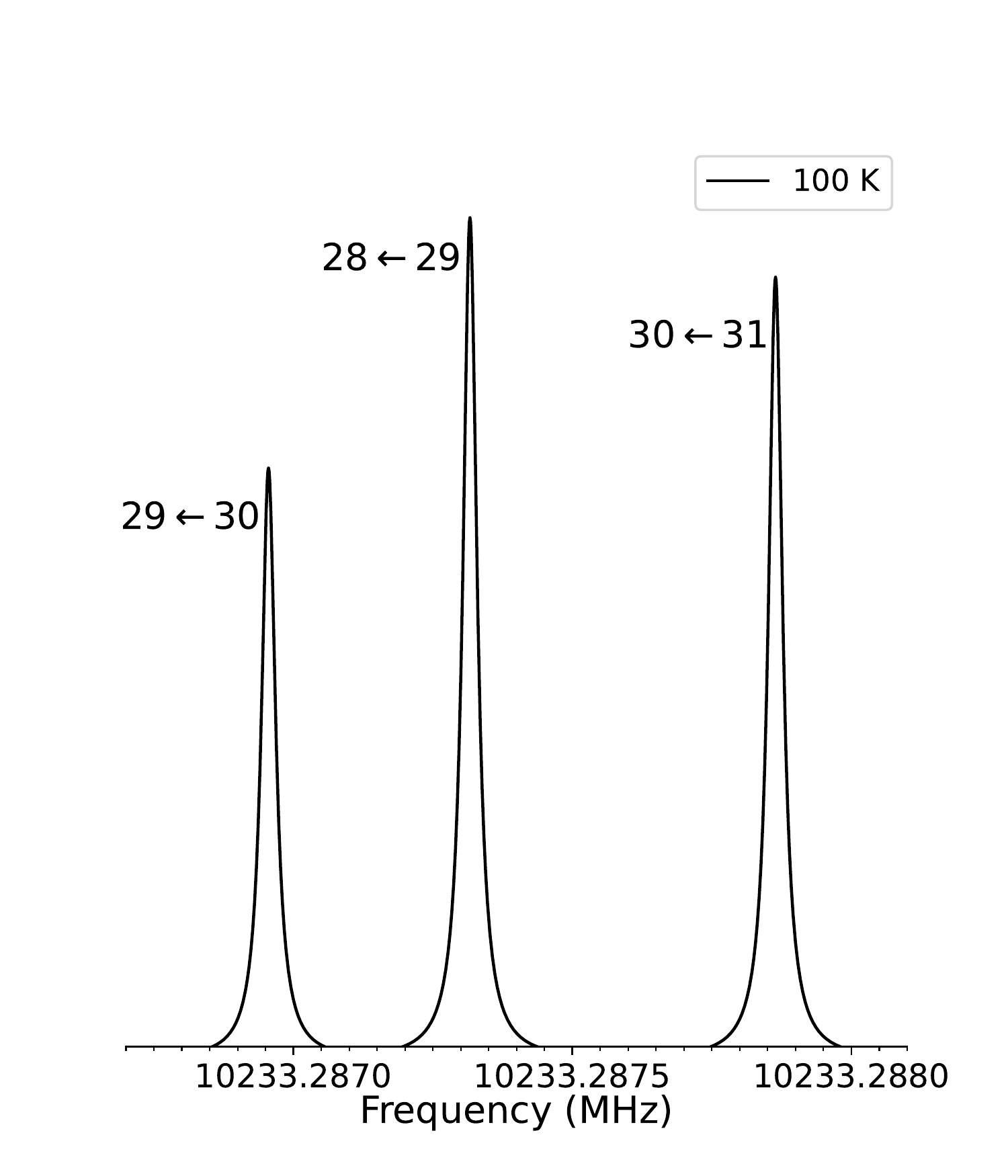}
         \caption{J$'_{K'_a,K'_c}$ $\rightarrow$ J"$_{K"_a,K"_c}$ = 30$_{0,30}$ $\rightarrow$ 29$_{1,29}$}
         \label{}
     \end{subfigure}  
     \caption{Same as Figure 6 for N-coronene.}
\end{figure*}
\begin{figure*}
     \centering
         \begin{subfigure}[b]{0.331\textwidth}
         \centering
         \includegraphics[width=\textwidth]{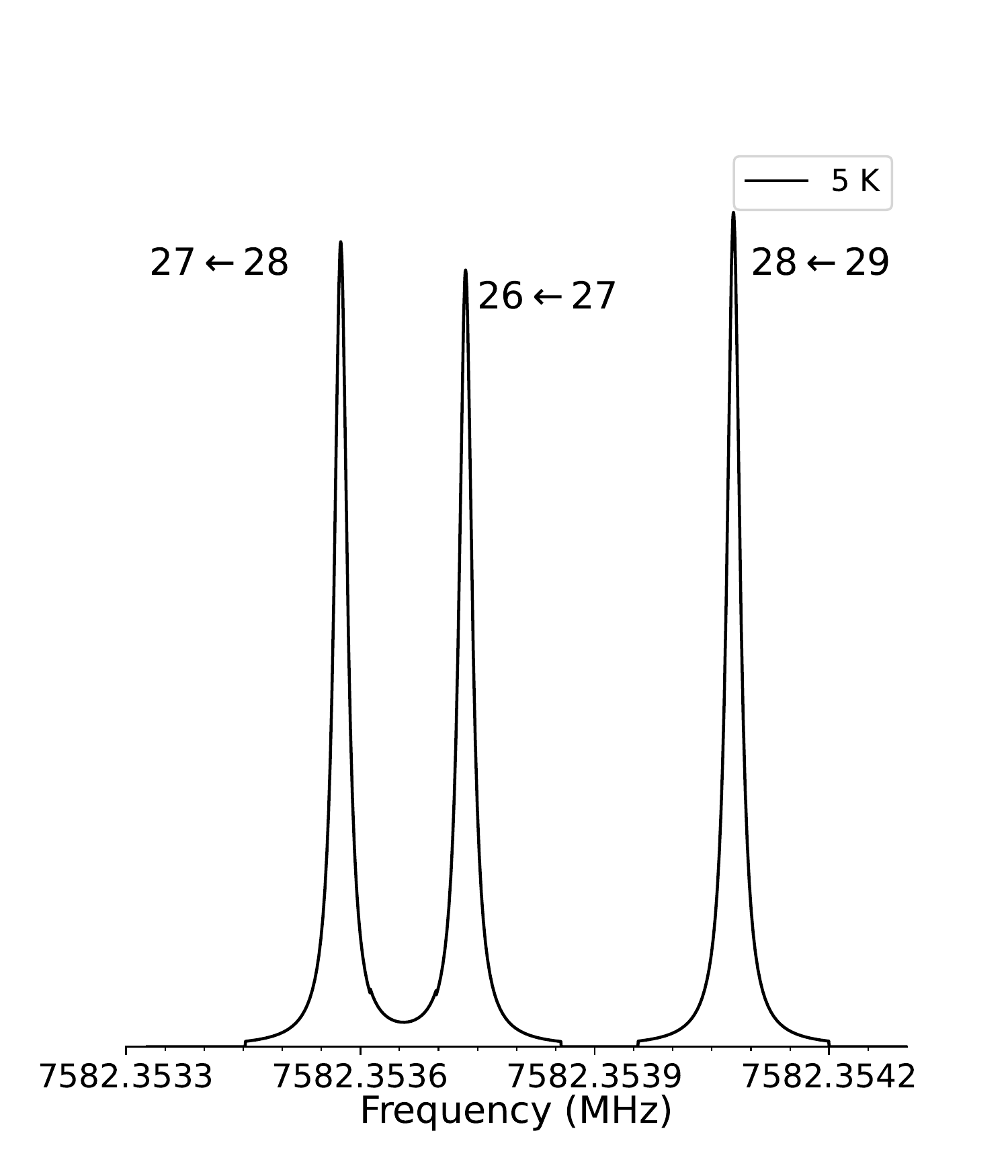}
         \caption{J$'_{K'_a,K'_c}$ $\rightarrow$ J"$_{K"_a,K"_c}$ = 28$_{0,28}$ $\rightarrow$ 27$_{0,27}$}
         \label{}
     \end{subfigure}
         \begin{subfigure}[b]{0.331\textwidth}
         \centering
         \includegraphics[width=\textwidth]{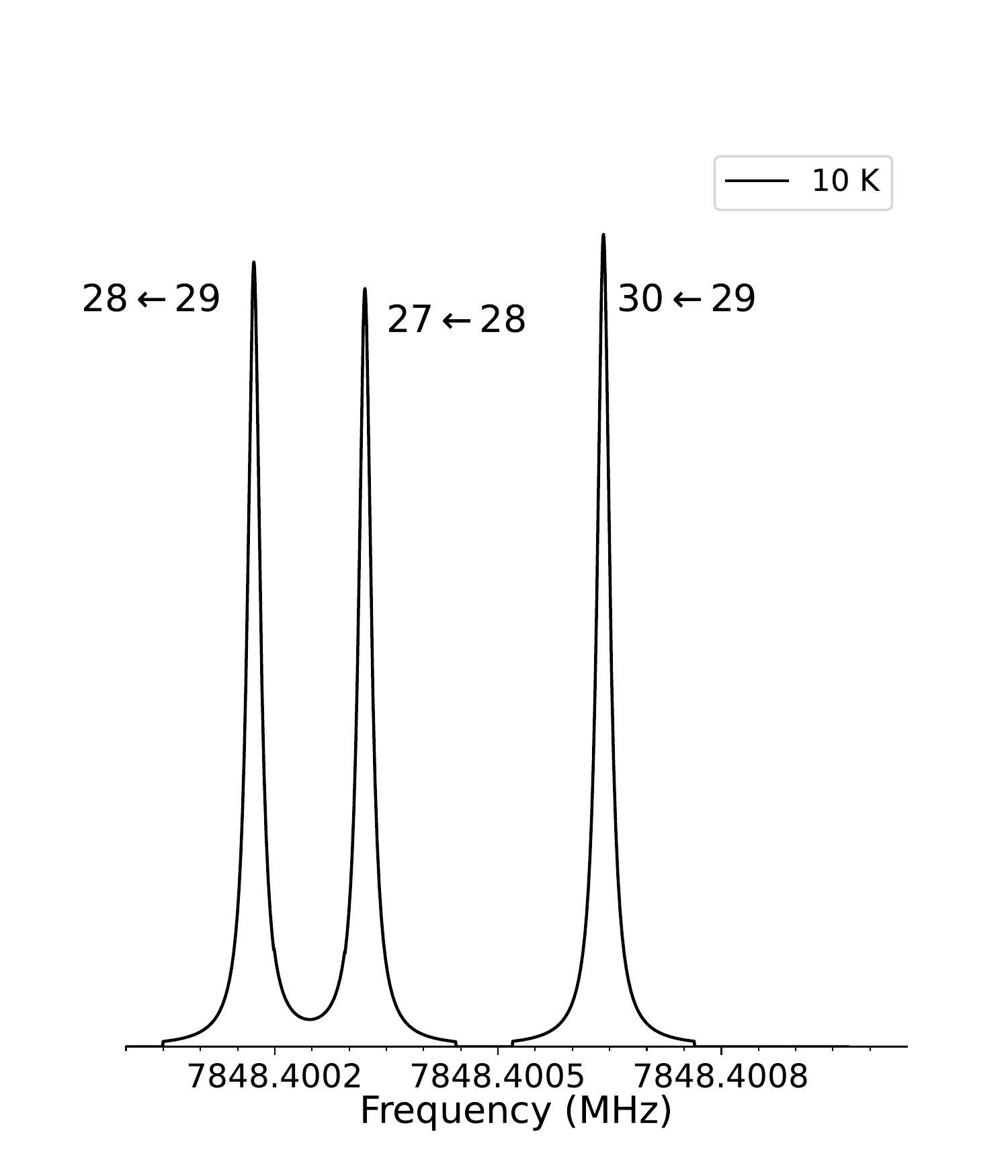}
         \caption{J$'_{K'_a,K'_c}$ $\rightarrow$ J"$_{K"_a,K"_c}$ = 29$_{0,29}$ $\rightarrow$ 28$_{1,28}$}
         \label{}
     \end{subfigure}  
         \begin{subfigure}[b]{0.331\textwidth}
         \centering
         \includegraphics[width=\textwidth]{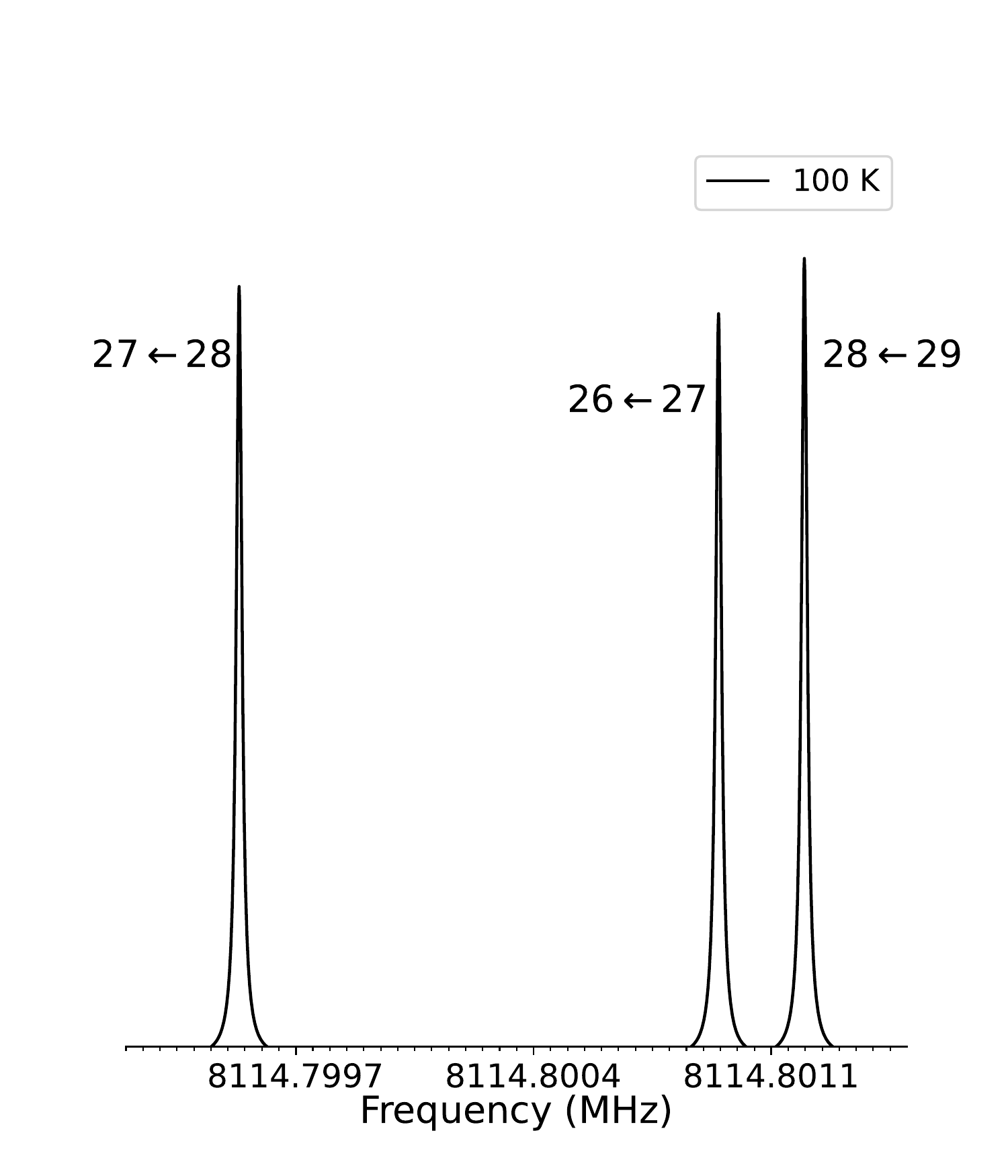}
         \caption{J$'_{K'_a,K'_c}$ $\rightarrow$ J"$_{K"_a,K"_c}$ = 28$_{3,26}$ $\rightarrow$ 27$_{2,25}$}
         \label{}
     \end{subfigure}  
     \caption{Same as Figure 6 for CN-coronene.}
\end{figure*}
\section{astrophysical implications and conclusions}
In radio astronomy, the strength of a molecule's permanent dipole moment is a major factor in its ability to be detected in the ISM. Interstellar molecules are mostly discovered in space at radio frequencies using pure rotational spectroscopy. The CN-PAHs are found to be far more polar compared to the N-PAHs, with average dipole moment values of 5.5 and 2.7 D for the CN-PAHs and N-PAHs, respectively. The search of N- and CN-PAHs with radio telescopes has been undertaken towards cold molecular clouds, such as Tauras molecular cloud (TMC-1) \citep{2018Sci...359..202M,2021Sci...371.1265M} and Perseus molecular cloud \citep{2015MNRAS.447..315A}. In these clouds, the rotational transitions of PAH-related species are thought to be the strongest in the 8--33.5 GHz frequency range with typical temperature range of 5--10 K \citep{2020ApJ...900L..10M}. The only PAH species discovered so far in the ISM are CN-PAHs; 1- and 2-cyanonaphthalene \citep{2021Sci...371.1265M}.  

We report the cm-wave spectra of N- and CN-pyrene (C$_{16}$H$_{10}$) and N- and CN-coronene (C$_{24}$H$_{12}$) for the first time. The B3LYP/6-311+G(d,p) level of theory computes the rotational constants within $\sim$0.6 per cent of those obtained experimentally for all the N- and CN-PAHs used here as test molecules. For rotational transition frequencies, the B3LYP/6-311+G(d,p) level achieves best performance here and shows relative errors of $\sim$0.5 and $\sim$0.18 per cent between experimental and theoretical values for N-PAH and CN-PAHs, respectively. As a result, the same level of theory has also been utilized for computing the molecular constants and rotational spectra of pyrene and coronene in N-PAH and CN-PAH variants. The data presented here may be directly utilized to detect these PAHs in potential ISM environments. 

While taking interstellar detection into the consideration, the hyperfine resolved structure might play a significant role for the considered PANH species as the splitting in the cm-wave region is found to be similar with the linewidths measured for a typical cold, dark molecular clouds-- e.g., TMC-1 \citep{2021JMoSp.37711425B}. Therefore, the $^{14}$N quadrupole hyperfine resolved structures are also given for the strongest transitions of the considered N- and CN-PAH molecules at 5, 10 and 100 K. 
\subsubsection{Pyrene}
\begin{table}\small
\begin{threeparttable}
	\centering
	\caption{Hyperfine resolved strongest rotational transitions of pyrene and coronene in N- and CN-PAH variants.}
	\label{tab:example_table}
	\begin{tabular}{lccc} 
		\hline
		\hline
		\multicolumn{2}{c}{Transition} &
		\multicolumn{1}{c}{Frequency} & T\\
		\cmidrule(lr){1-2}

		J$'_{K'_a,K'_c}$ $\rightarrow$ J$"_{K"_a,K"_c}$ & $F' \rightarrow F"$ & & \\ 
		 & & (MHz) & (K)\\
		\hline
	\hspace{-0.25cm} N-pyrene &&&\\	
	$23_{1,23} \rightarrow 22_{0,22}$	& 23 $\rightarrow$ 22 & 17009.8637 & 5\\
	 & 22 $\rightarrow$ 21 & 17009.8641 & \\
		& 24 $\rightarrow$ 23 & 17009.8657  &\\
	$26_{1,26} \rightarrow 25_{0,25}$	& 26 $\rightarrow$ 25 & 19181.3069 & 10\\
	 & 25 $\rightarrow$ 24 & 19181.3073 & \\
		& 27 $\rightarrow$ 26 & 19181.3085  &\\
	$28_{3,26} \rightarrow 27_{2,25}$	& 28 $\rightarrow$ 27 & 22078.7859 & 100\\
	 & 27 $\rightarrow$ 26 & 22078.7892 & \\
		& 29 $\rightarrow$ 28 & 22078.7901  &\\\hline
	\hspace{-0.25cm} CN-pyrene &&&\\
	$28_{1,28} \rightarrow 27_{0,27}$	& 28 $\rightarrow$ 27 & 14703.8466 & 5\\
	 & 27 $\rightarrow$ 26 & 14703.8468 & \\
		& 29 $\rightarrow$ 28 & 14703.8487  &\\
	$29_{1,29} \rightarrow 28_{0,28}$	& 29 $\rightarrow$ 28 & 15219.7623 & 10\\
	 & 28 $\rightarrow$ 27 & 15219.7624 & \\
		& 30 $\rightarrow$ 29 & 15219.7643  &\\
	$29_{2,28} \rightarrow 28_{1,27}$	& 29 $\rightarrow$ 28 & 15736.3680 & 100\\
	 & 28 $\rightarrow$ 27 & 15736.3693 & \\
		& 30 $\rightarrow$ 29 & 15736.3702  &\\\hline
	\hspace{-0.25cm} N-coronene &&&\\
	$28_{1,28} \rightarrow 27_{1,27}$	& 29 $\rightarrow$ 28 & 9562.2517 & 5\\
	 & 28 $\rightarrow$ 27 & 9562.2522 & \\
		& 30 $\rightarrow$ 29 & 9562.2527  &\\
	$29_{0,29} \rightarrow 28_{0,28}$	& 30 $\rightarrow$ 29 & 9897.7693 & 10\\
	 & 29 $\rightarrow$ 28 & 9897.7697 & \\
		& 31 $\rightarrow$ 30 & 9897.7702  &\\
	$30_{1,30} \rightarrow 29_{1,29}$	& 30 $\rightarrow$ 29 & 10233.2869 & 100\\
	 & 29 $\rightarrow$ 28 & 10233.2873 & \\
		& 31 $\rightarrow$ 30 & 10233.2878  &\\\hline
	\hspace{-0.25cm} CN-coronene &&&\\
	$28_{0,28} \rightarrow 27_{1,27}$	& 28 $\rightarrow$ 27 & 7582.3535 & 5\\
	 & 27 $\rightarrow$ 26 & 7582.3537 & \\
		& 29 $\rightarrow$ 28 & 7582.3549  &\\
	$29_{0,29} \rightarrow 28_{1,28}$	& 29 $\rightarrow$ 28 & 7848.4001 & 10\\
	 & 28 $\rightarrow$ 27 & 7848.4003 & \\
		& 30 $\rightarrow$ 29 & 7848.4006  &\\
	$28_{3,26} \rightarrow 27_{2,25}$	& 28 $\rightarrow$ 27 & 8114.7995 & 100\\
	 & 27 $\rightarrow$ 26 & 8114.8010 & \\
		& 29 $\rightarrow$ 28 & 8114.8011  &\\
	\hline
    \end{tabular}
    \end{threeparttable}
    \end{table}
The hyperfine resolved structures are shown at 5, 10 and 100 K in Figure 6 and Figure 7 for N- and CN-pyrene, respectively while the transition frequencies are tabulated in Table 4. The complete rotational spectra at 10, 100 and 300 K are given in Figure A1. For both N- and CN-pyrene, the strongest lines lie at cm-wavelengths at low as well as some warmer temperatures. As temperature increases, the larger partition function significantly reduces the strength of transitions (Figure A1). For N-pyrene: at 5 K, the strongest transition is J'=23 around 17 GHz falling in K$_{\rm u}$ band whereas at 10 K and 100 K, the strongest transitions lie within K band, corresponding to J'=26 and J'=28 near 19.1 and 22 GHz, respectively (Fig. 6 and Table 4).    

For CN-pyrene: at 5, 10 and 100 K, the strongest lines fall in K band. The strongest transition at 5 K is J'=28 near 14 GHz while at 10 and 100 K, the strongest lines correspond to J'=29 around 15.2 and 15.7 GHz, respectively (Fig. 7 and Table 4). 
\subsubsection{Coronene}
For coronene, Figure 8 and Figure 9 show hyperfine splitting of the strongest rotational transition of N- and CN-coronene, respectively at 5, 10 and 100 K, which are tabulated in Table 4. The corresponding rotational spectra in the complete wavelength range are shown in Figure A1. The strongest lines in coronene are present at shorter wavelengths compared to pyrene at each temperature (Figure A1). In terms of the intensity, the transitions in coronene are weaker than those in pyrene for both the N- and CN-PAH variants (Figure A1) due to its larger rotational partition function that depends on rotational constants (A, B and C) (Q$_{\rm rot}$ $\propto$ $\sqrt{\frac{\rm T^{3}}{\rm ABC}}$, where T is temperature). 

The strongest lines of N-coronene fall in X band (8--12 GHz) at cold and some warmer temperatures (Figure 8). The strongest transitions of N-coronene at 5, 10 and 100 K are J'=28, J'=29 and J'=30 around 9.5, 9.8 and 10.2 GHz, respectively. For CN-coronene, the wavelengths are shorter than those of N-coronene and this behavior is consistent with that of pyrene (Table 4). The strongest transitions of CN-coronene at 5, 10 and 100 K correspond to J'=28, J'=29 and J'=28 near 7.5, 7.8 and 8.1 GHz, respectively. 

Based on the large permanent dipole moment, CN-PAHs are the most suitable PAH variant to be searched for in the ISM. Furthermore, the smaller partition function in pyrene renders CN-pyrene an excellent candidate to be found in cold, dark molecular clouds such as the TMC-1. The search for CN-pyrene in molecular clouds like TMC-1 is supported by the following arguments.

It is believed that the PAHs form in carbon-rich circumstellar environments (CSEs), where there is a significant elemental abundance of nitrogen \citep{1999ApJ...524..831R}. As a result, the PAH family members that are ejected into the ISM from CSEs and PNe include a mixture of pure PAHs, CN-PAHs, and N-PAHs \citep{2021ApJ...917L..35A}. Pure PAH lacks the permanent dipole moment and does not show a pure rotational spectrum. For N-containing PAHs, the detection through rotational lines might be hampered in the regions exhibiting the strong 6.2 $\mu$m band due to the following reasons. Large-sized PANHs show smaller rotational constants thus possess larger partition functions \citep{2005ApJ...632..316H,2008ARA&A..46..289T} and PAHs with more than $\sim$40--50 C atoms are thought to be more relevant to be found in such regions. Their radiated power might be dispersed over many weak lines, which makes their detection quite challenging \citep{2014MNRAS.437.2728A}.

The detection of CN-PAHs in TMC-1 is remarkable because it appears that they were formed in the cold, dark molecular cloud itself. It is plausible that alternative interstellar PAH formation mechanisms and chemistry exist in such molecular clouds because the detected PAHs are too small to persist in the diffuse ISM \citep{2021Sci...371.1265M,2021ApJ...917L..35A}. The present work may also be crucial for guiding laboratory experiments and inspiring future experimental work on the rotational spectroscopy of CN-PAHs and N-PAHs.

\section*{Acknowledgements}
We thank to the anonymous reviewer for the comments and suggestions that
helped in bringing clarity to this work. AV acknowledges research fellowship from DST SERB (SERB-EMR/2016/005266) and UGC. AP acknowledges financial support from DST SERB (SERB-CRG/2021/000907), IoE incentive grant, BHU (incentive/2021-22/32439), Banaras Hindu University, Varanasi and thanks the Inter-University Centre for Astronomy and Astrophysics, Pune for associateship. The authors also acknowledge support from DST JSPS grant (DST/INT/JSPS/P-238/2017).
\section*{Data availability}
The data used in this article will be shared upon reasonable request to corresponding author.
\bibliographystyle{mnras}
\bibliography{sample63}
\appendix
\section{Calculated rotational spectra at different temperatures}
\begin{figure*}
     \centering
         \begin{subfigure}[b]{0.45\textwidth}
         \centering
         \includegraphics[width=\textwidth]{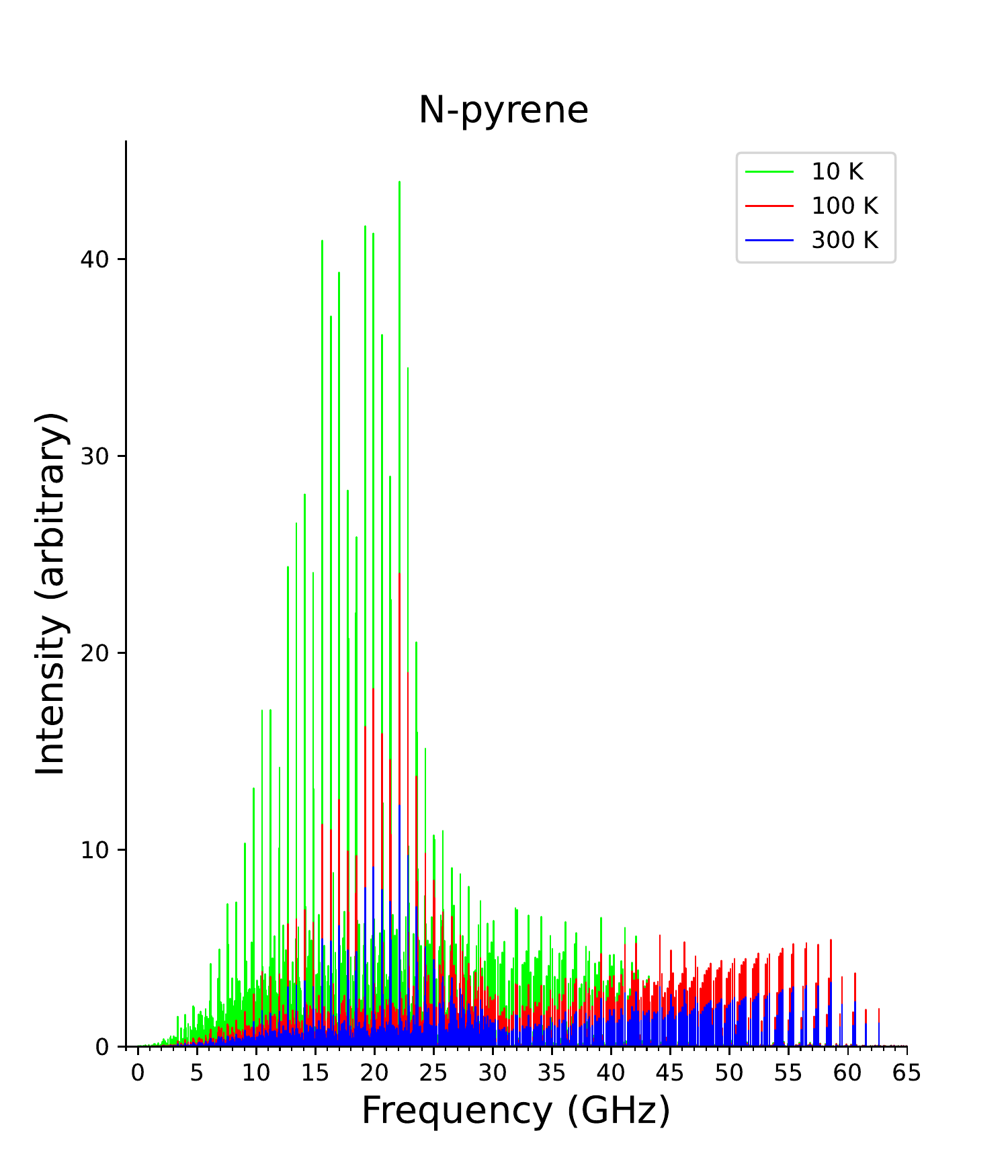}
         \caption{}
         \label{}
     \end{subfigure}
         \begin{subfigure}[b]{0.45\textwidth}
         \centering
         \includegraphics[width=\textwidth]{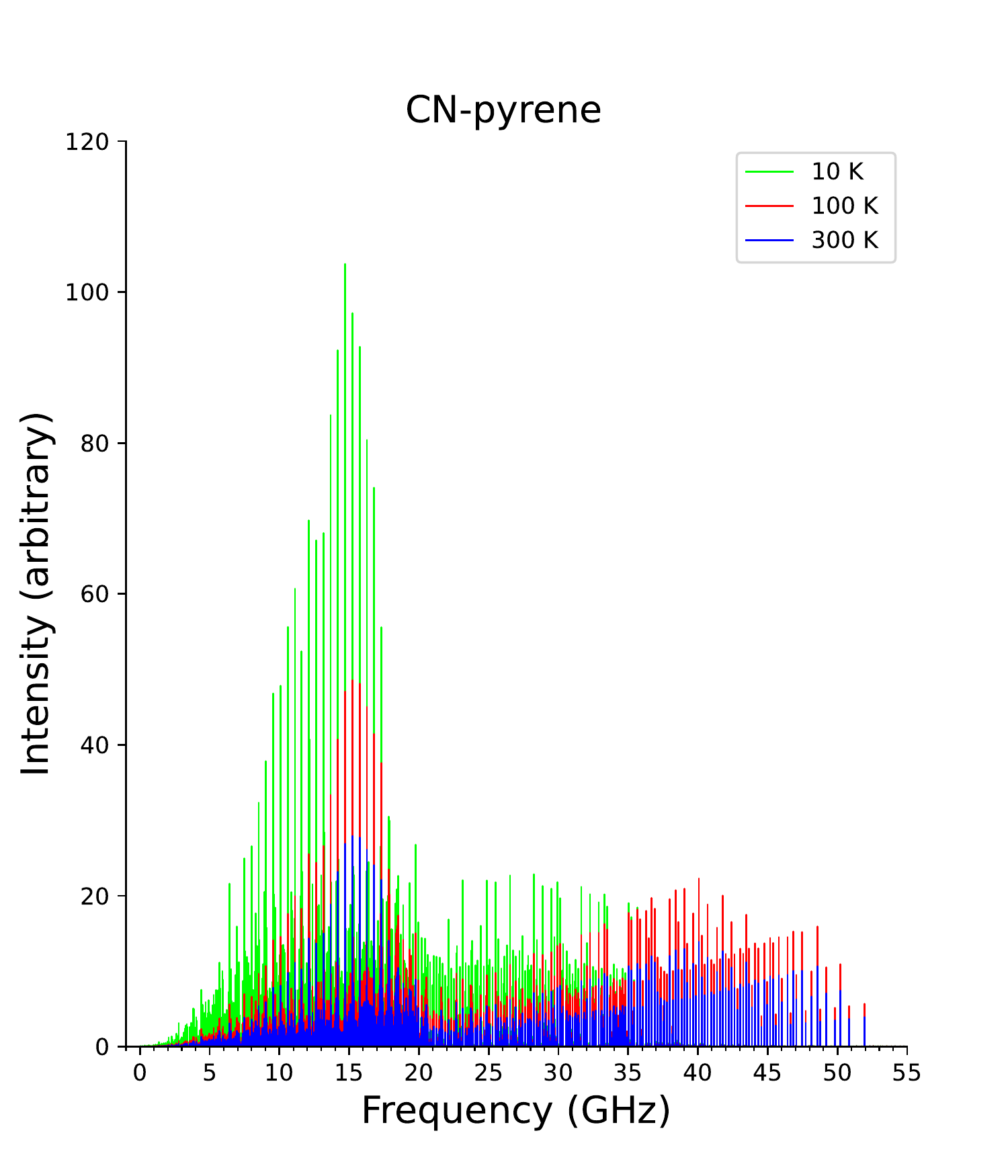}
         \caption{}
         \label{}
     \end{subfigure}  
         \begin{subfigure}[b]{0.45\textwidth}
         \centering
         \includegraphics[width=\textwidth]{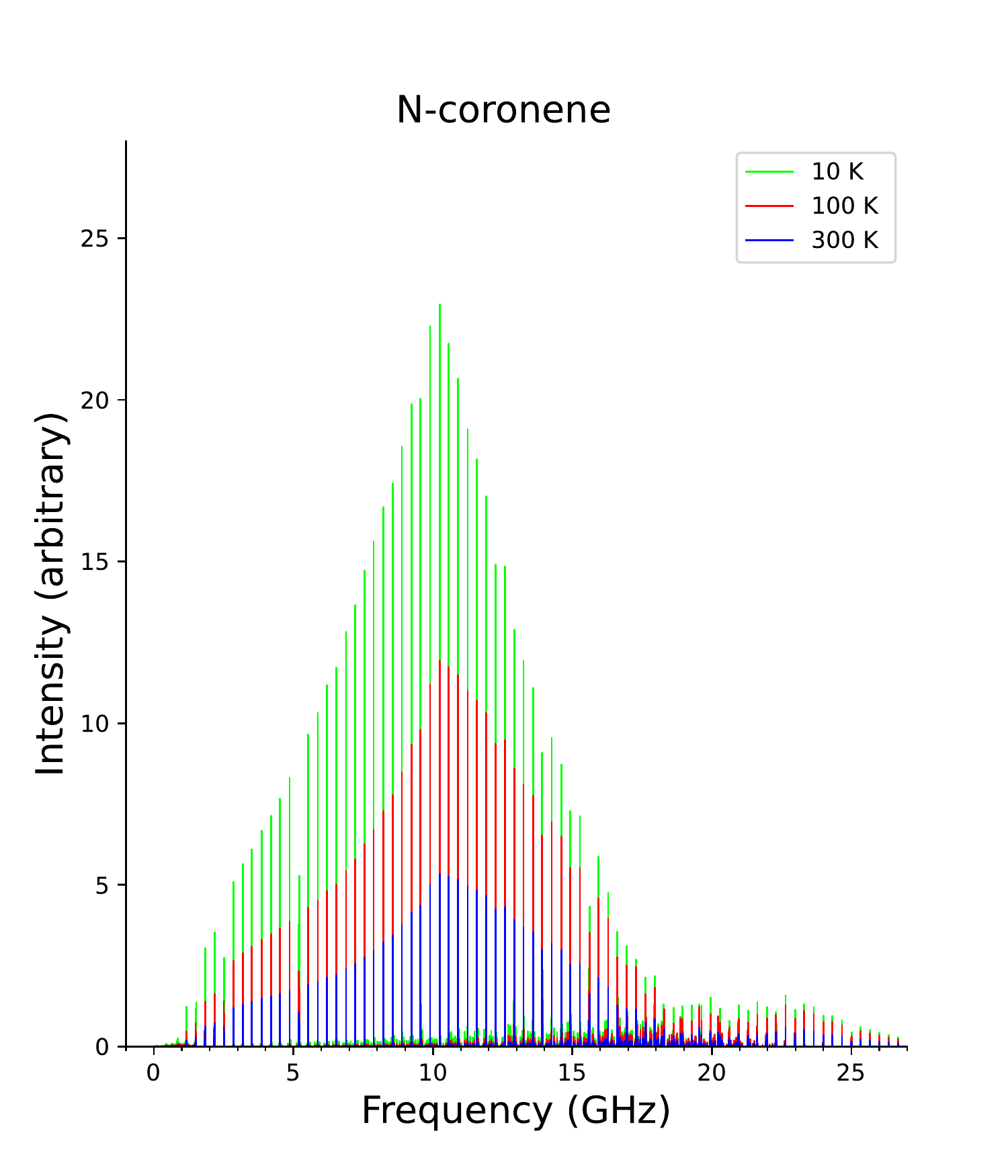}
         \caption{}
         \label{}
     \end{subfigure}  
         \begin{subfigure}[b]{0.45\textwidth}
         \centering
         \includegraphics[width=\textwidth]{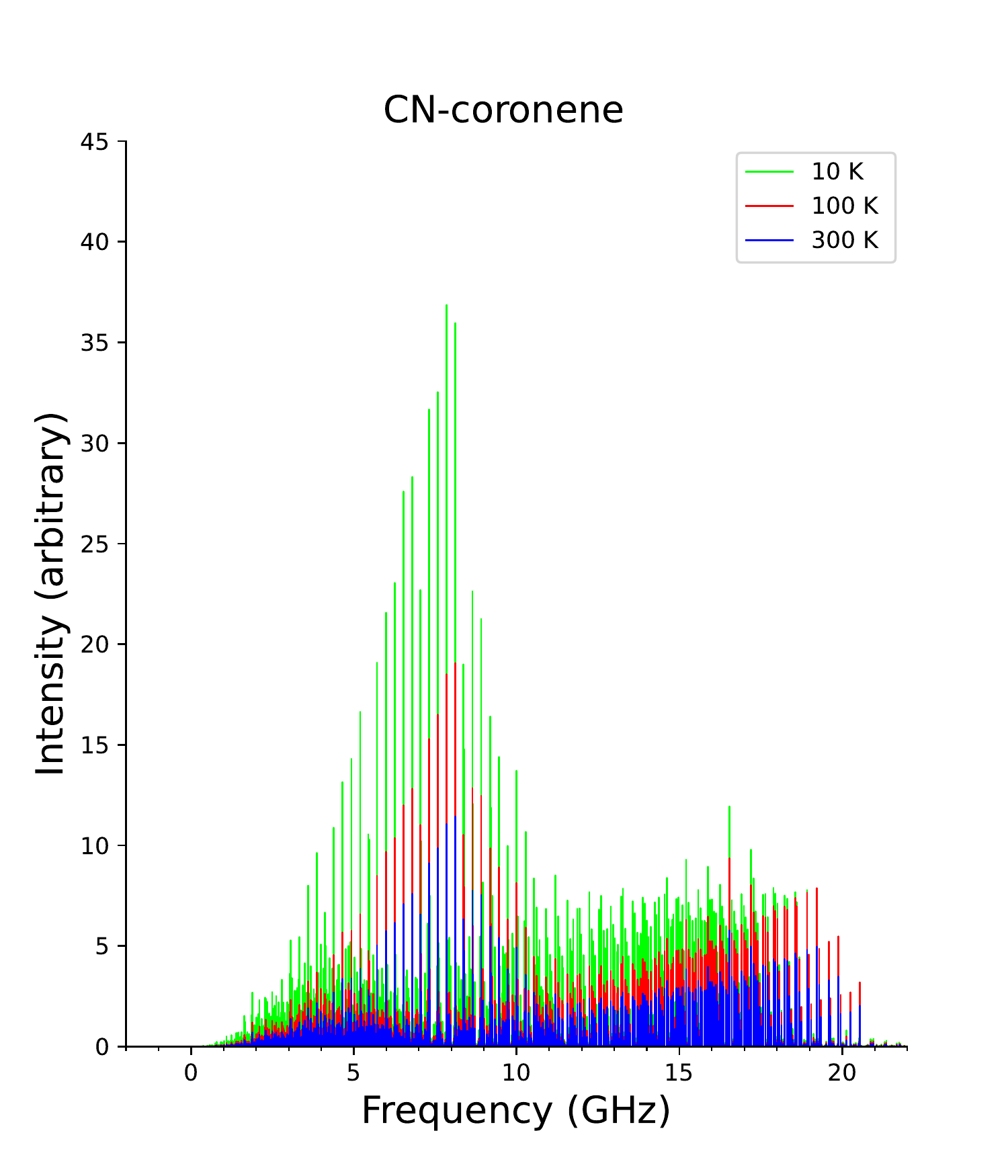}
         \caption{}
         \label{}
     \end{subfigure}  
     \caption{Calculated rotational spectra of the considered N- and CN-PAHs at 10 K (green), 100 K (red) and 300 K (blue).}
\end{figure*}
\bsp	
\label{lastpage}
\end{document}